\documentstyle[preprint,aps,floats,epsf]{revtex}

\def\gsim{\, \lower0.5ex\hbox{$\stackrel{>}{\sim}$}\, }
\def\lsim{\, \lower0.5ex\hbox{$\stackrel{<}{\sim}$}\, }

\def\figHadXsecA
{
\begin{figure}[hbtp]
 \epsfxsize=\hsize
 \centerline{\epsfbox{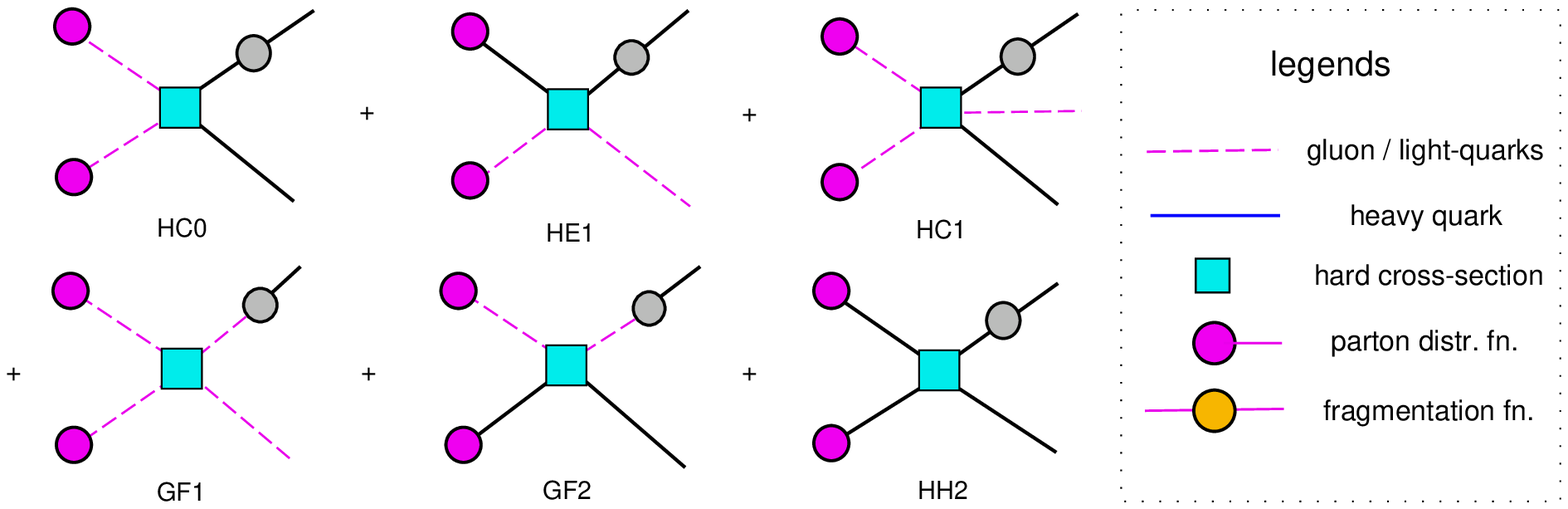}}
 \tightenlines
 \caption{
 Graphical representation of the leading terms in the factorization
 formula which correspond to the various production mechanisms.
 The initial state hadron line for the parton distributions are uniformly 
 suppressed. }
 \label{fig:HadXsecA}
\end{figure}
}

\def\figNloXsec
{
\begin{figure}[hbtp]
 \epsfxsize=5.8 in
 \centerline{\epsfbox{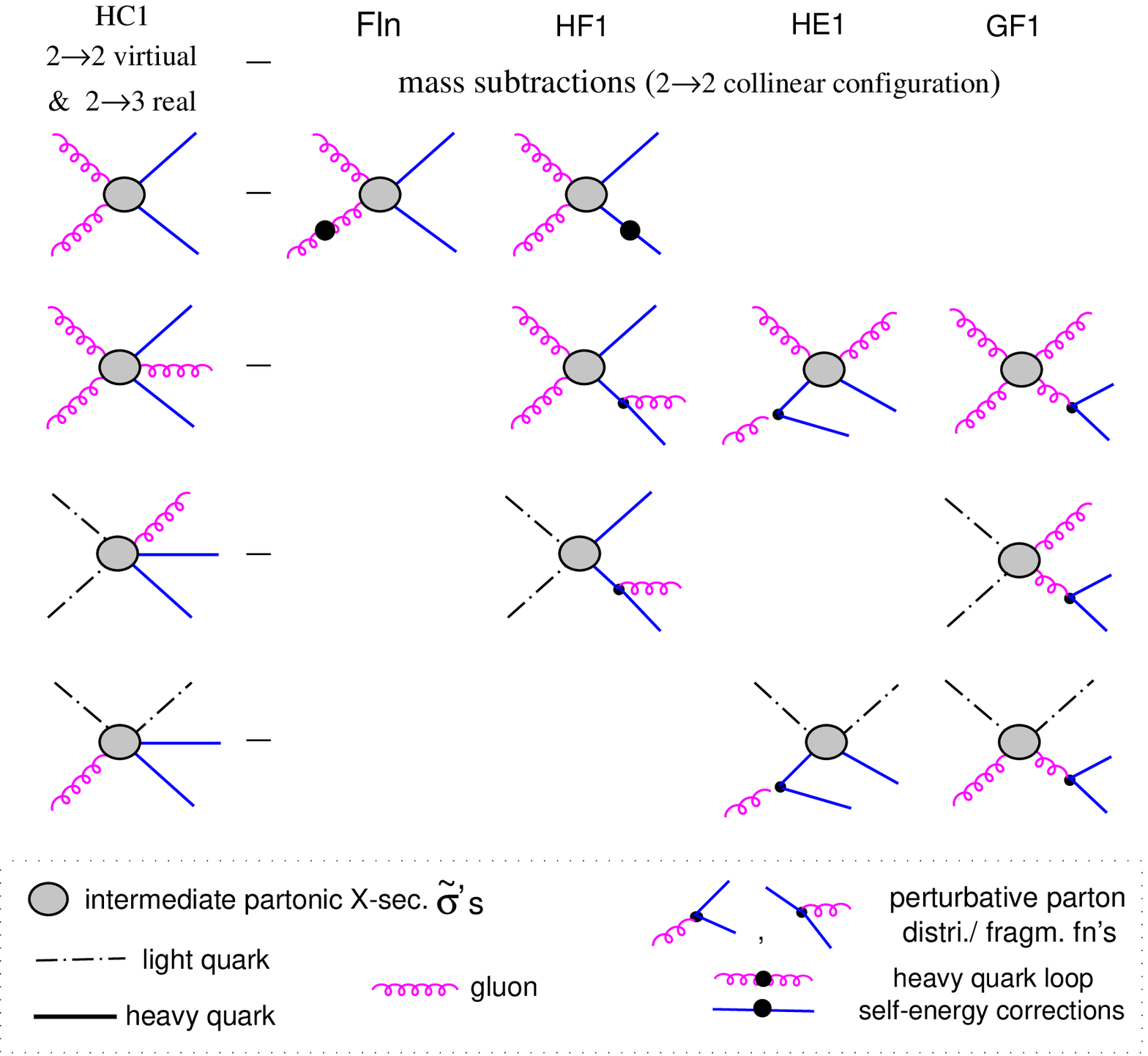}}
 \tightenlines
 \caption{
 Graphical representation of the terms in the right-hand-side
 of Eq.~{\protect \ref{nloIRSxsec}}.
 Collinear singularities due to light partons have already been
 subtracted.
 The vertices represented by a dot are the heavy quark parts of
 the perturbative distribution and fragmentation functions,
 as in Eqs.\
 \protect\ref{pertpdf} and \protect\ref{pertfrg}.
 }
 \label{fig:NloXsec}
\end{figure}
}

\def\figHadXsecB
{
\begin{figure}[hbtp]
 \epsfysize=7.5 in
  \centerline{\epsfbox{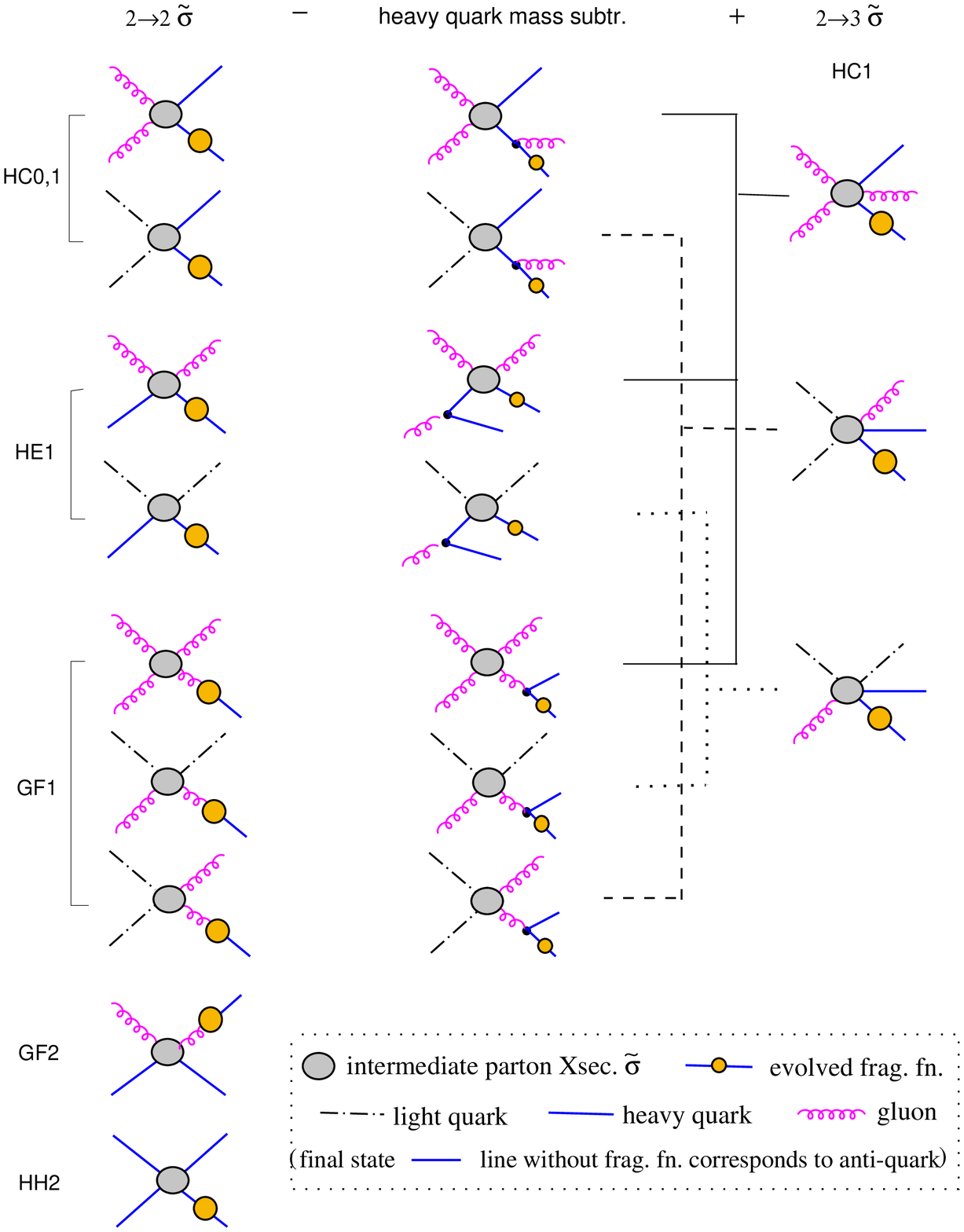}}
 \tightenlines
 \caption{
 Graphical representation of the physical
cross-section, Eq.~{\protect \ref{HadXsec1}} and
Fig.~{\protect \ref{fig:HadXsecA}}, written in terms of the intermediate
partonic cross-sections $^n\tilde{\sigma}$'s and the attendant
heavy quark mass subtraction terms which represent the overlap between
the 2$\rightarrow$2 and 2$\rightarrow$3 cross-sections. Initial state parton
distribution function factors are uniformly suppressed.}
 \label{fig:HadXsecB}
\end{figure}
}

\def\figCompensate
{
\begin{figure}[hbtp]
 \begin{center}
 \leavevmode
 \epsfxsize=0.4\hsize
 \epsfbox{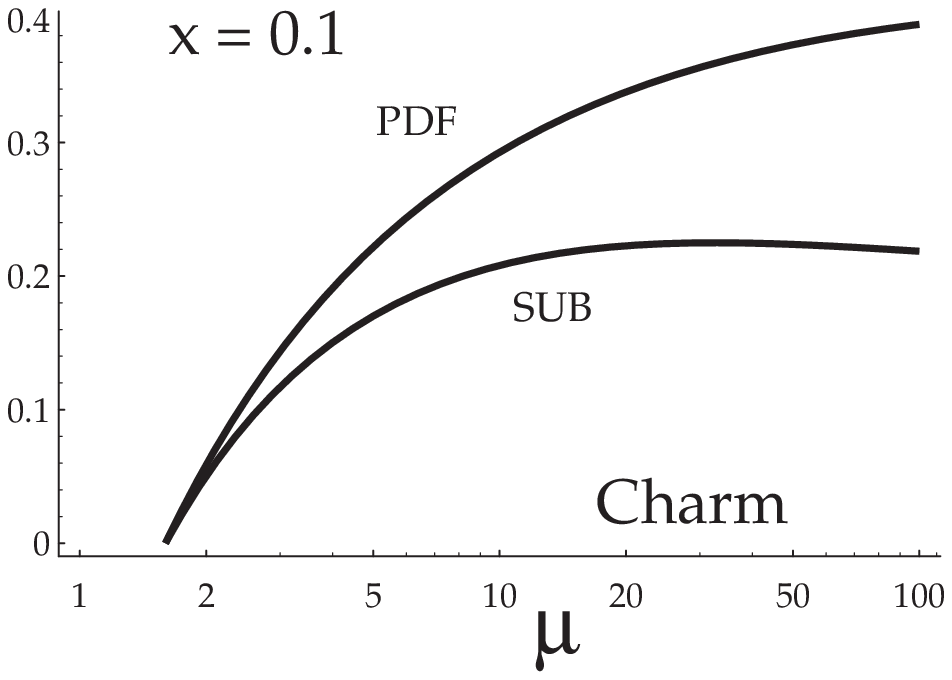}
 \hfill
 \epsfxsize=0.4\hsize
 \epsfbox{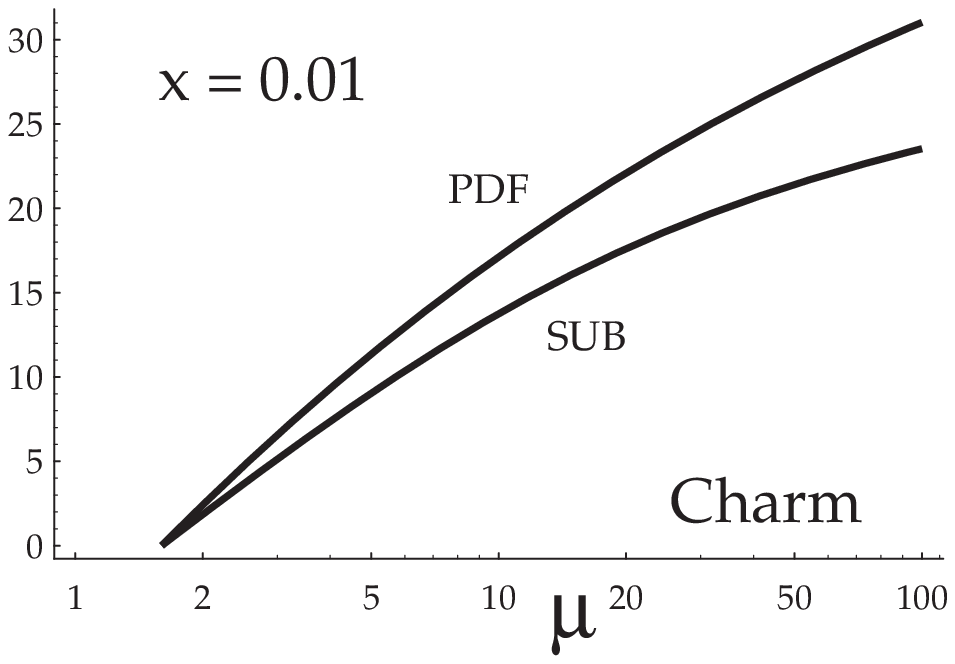}\vspace*{0.5cm}
 \epsfxsize=0.4\hsize
 \epsfbox{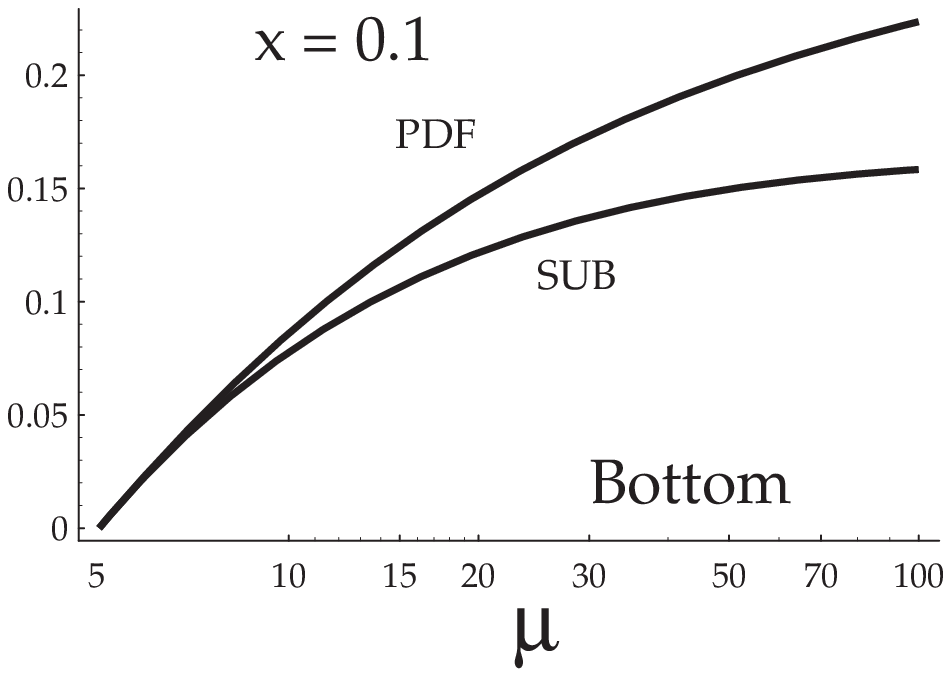}
 \hfill
 \epsfxsize=0.4\hsize
 \epsfbox{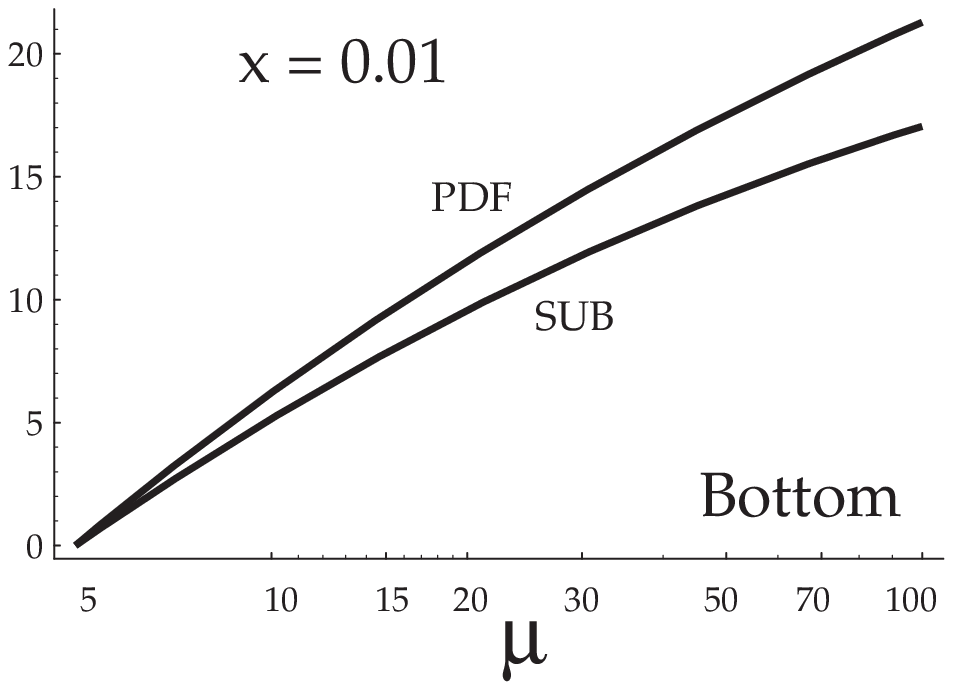}
 \end{center}
 \tightenlines
 \caption{
 Comparison of the evolved PDFs, $f^H(x,\mu)$ (labeled PDF), and
 perturbative PDFs, $^1f^H(x,\mu)$ (labeled SUB),
 as a function of the renormalization scale $\mu$
 for charm at $x=0.1$ (a) and $x=0.01$ (b), and
 for bottom at $x=0.1$ (c) and $x=0.01$ (d). This shows the
 compensation between fully evolved heavy quark
 parton distribution and the first order perturbative contribution 
(which is the only part contained in the FFN scheme calculation).
 }
 \label{fig:Compensate}
\end{figure}
}

\def\figonex
{
\begin{figure}[hbtp]
 \epsfxsize=\hsize
 \centerline{\epsfbox{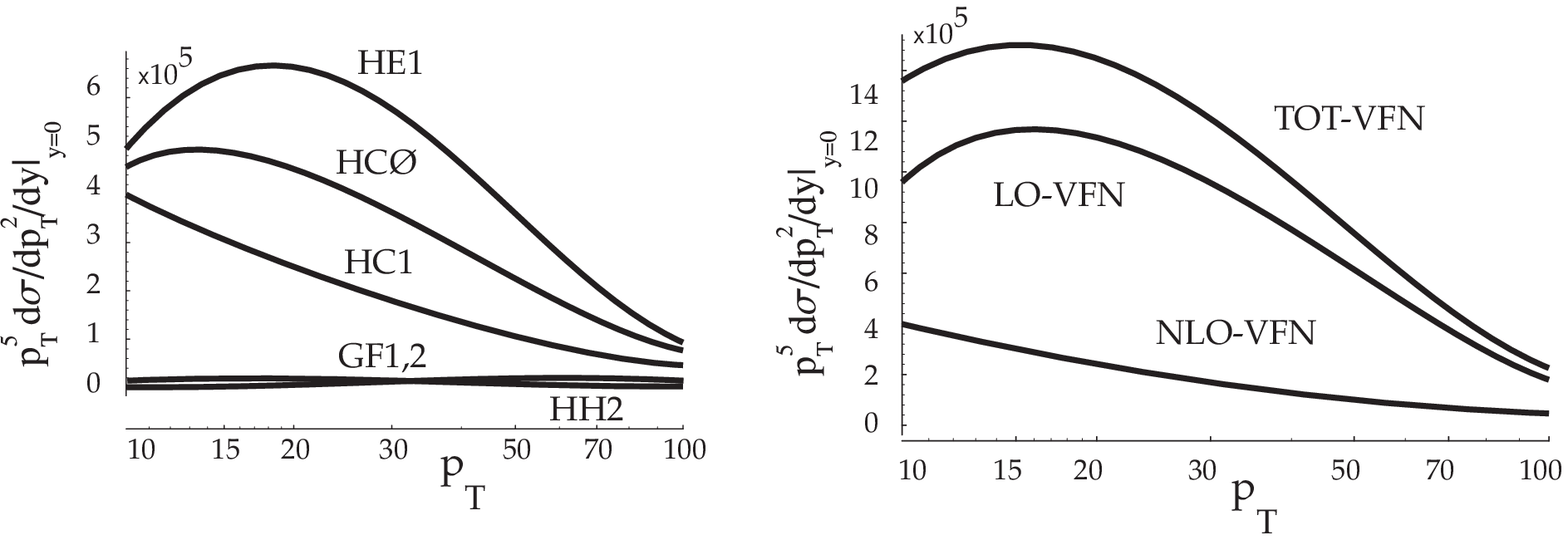}}
 \tightenlines
 \caption{
 Contributions to the scaled cross section
 $p_t^5 \, d\sigma/dp_t^2 /dy|_{y=0}\   (nb\,  {\rm GeV}^{3})$
 vs.\ $p_t$ for $b$ production
 at 1800 GeV with $\mu=M_T/{\protect \sqrt{2}}$
 organized in the  ACOT formalism.
 a) The curves correspond to the separate terms of
 Eq.~{\protect \ref{HadXsec1}}.
 b) The curves are leading-order ($\alpha_s^2$) (LO-VFN),
 next-to-leading-order ($\alpha_s^3$) (NLO-VFN), 
and the total result  (TOT-VFN).
 }
 \label{fig:Figx1}
\end{figure}
}

\def\figtwox
{
\begin{figure}[hbtp]
 \epsfxsize=\hsize
 \centerline{\epsfbox{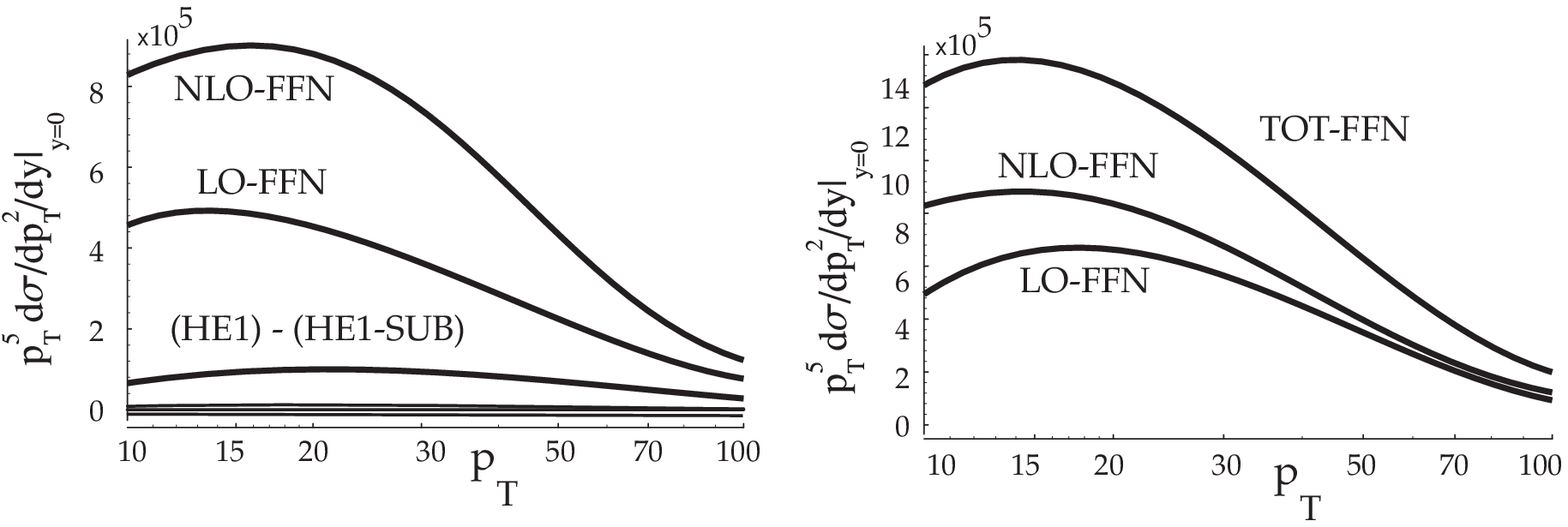}}
 \tightenlines
 \caption{
 Contributions to the scaled cross section
 $p_t^5 \, d\sigma/dp_t^2 /dy|_{y=0}\   (nb\,  {\rm GeV}^{3})$
 vs.\ $p_t$ for $b$ production
 at 1800 GeV with $\mu=M_T/{\protect \sqrt{2}}$
 organized according to the FFN scheme.
 a) The curves correspond to the order $\alpha_s^2$ (LO-FFN) and order 
$\alpha_s^3$ (NLO-FFN) heavy-flavor creation (HC) contributions (without heavy
mass subtractions). Also shown are the corrections due to the HE, GF1, GF2, 
and HH2 terms with associated subtractions as given in 
Fig.~{\protect \ref{fig:HadXsecB}}. The last three are numerically
negligible and appear at the bottom of the plot unlabelled.
 Note that, (i) the NLO term is two times {\em larger} then the LO one; 
 (ii) the contributions from  are small and unlabeled.
 b) LO-FFN and NLO-FFN contributions along with the total result (TOT-FFN). 
Cf.\ Fig.~{\protect \ref{fig:Figx1}} for comparison with the ACOT scheme 
case. 
 }
 \label{fig:Figx2}
\end{figure}
}

\def\figfourx
{
\begin{figure}[hbtp]
 \epsfxsize=0.5\hsize
 \centerline{\epsfbox{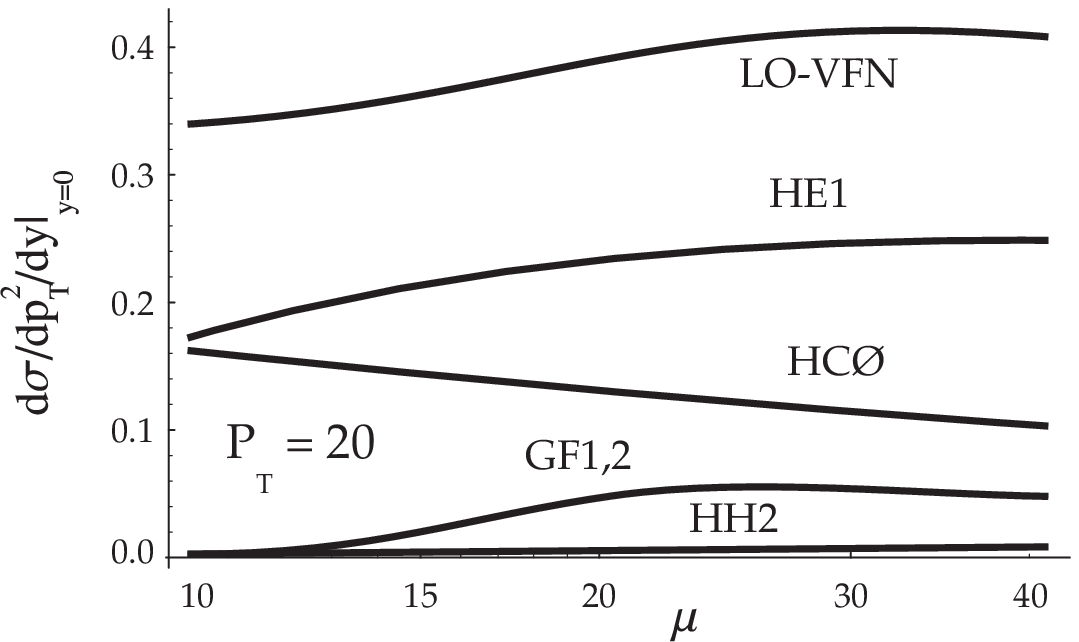}}
 \tightenlines
 \caption{
 Scale ($\mu$) dependence of the leading ($\alpha_s^2$) order contributions 
to the cross section
 $d\sigma/dp_t^2 /dy|_{y=0}\   (nb\,  {\rm GeV}^{-2})$ for $b$ production
 at 1800 GeV with $p_t=20$ GeV in the  ACOT formalism.
 }
 \label{fig:Figx4}
\end{figure}
}

\def\figfivex
{
\begin{figure}[hbtp]
 \epsfxsize=0.5\hsize
 \centerline{\epsfbox{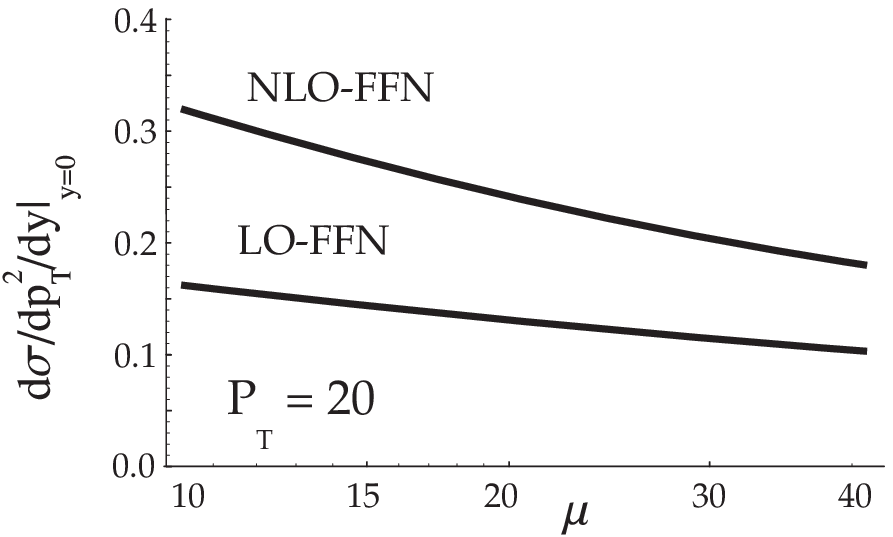}}
 \tightenlines
 \caption{
 Scale ($\mu$) dependence of the main contributions to the cross section
 $d\sigma/dp_t^2 /dy|_{y=0}\   (nb\,  {\rm GeV}^{-2})$
 for $b$ production
 at 1800 GeV with $p_t=20$ GeV organized in the  FFN scheme.
 }
 \label{fig:Figx5}
\end{figure}
}

\def\figsixx
{
\begin{figure}[hbtp]
 \epsfxsize=0.5\hsize
 \centerline{\epsfbox{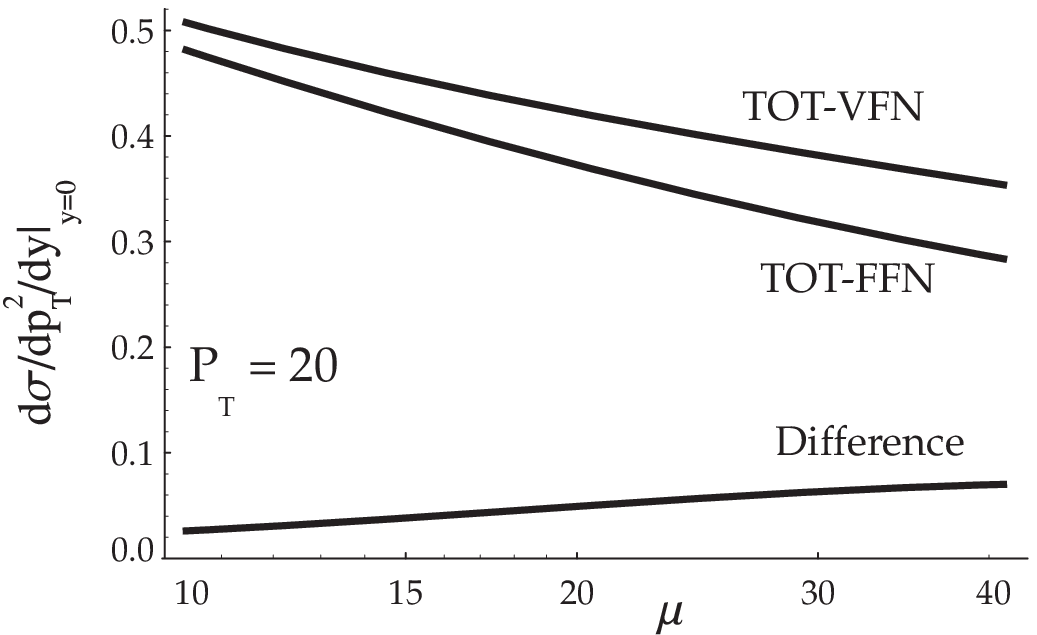}}
 \tightenlines
 \caption{
 Comparison of the total cross section 
 $d\sigma/dp_t^2 /dy|_{y=0}\   (nb\,  {\rm GeV}^{-2})$
 in the  ACOT and FFN formalism vs.\ $\mu$ for $b$ production
 at 1800 GeV with $p_t=20$ GeV.  The  {\em Difference} curve represents
 the additional resummed contributions included in the ACOT result.
 }
 \label{fig:Figx6}
\end{figure}
}

\def\figeightx
{
\begin{figure}[hbtp]
 \epsfxsize=0.5\hsize
 \centerline{\epsfbox{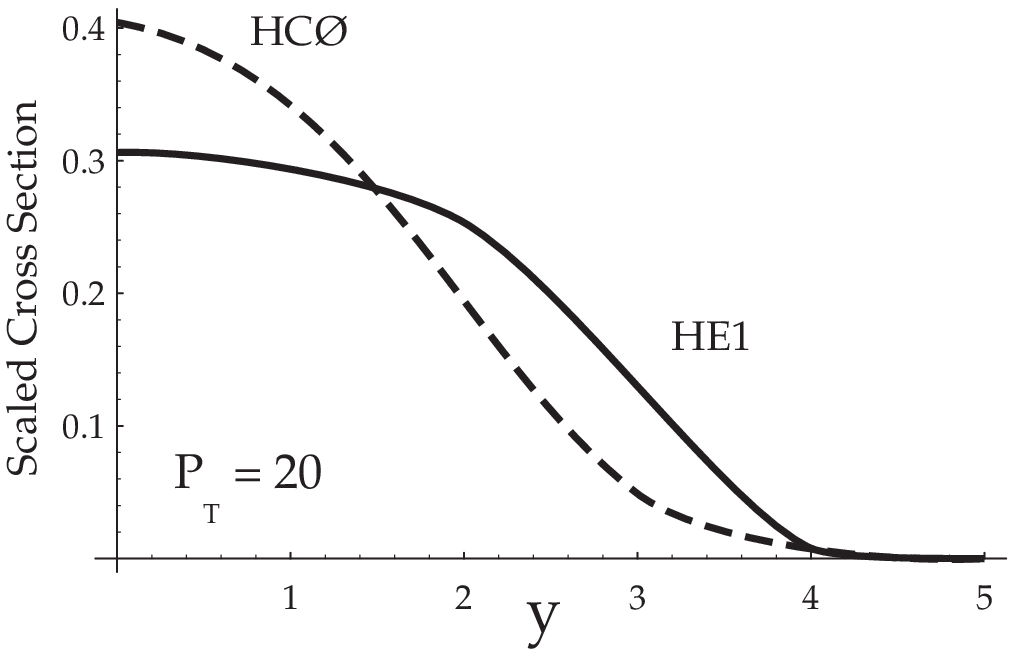}}
 \tightenlines
 \caption{
 Comparison of the rapidity ($y$) distribution for the HC\O\ and HE1 
 processes for $b$ production at 1800 GeV with $p_t=20$ GeV and $\mu = M_T$.  
 The
 curves are scaled to equal area to facilitate comparison of the shapes.
 }
 \label{fig:Figx8}
\end{figure}
}

\def\figninex
{
\begin{figure}[hbtp]
 \epsfxsize=0.5\hsize
 \centerline{\epsfbox{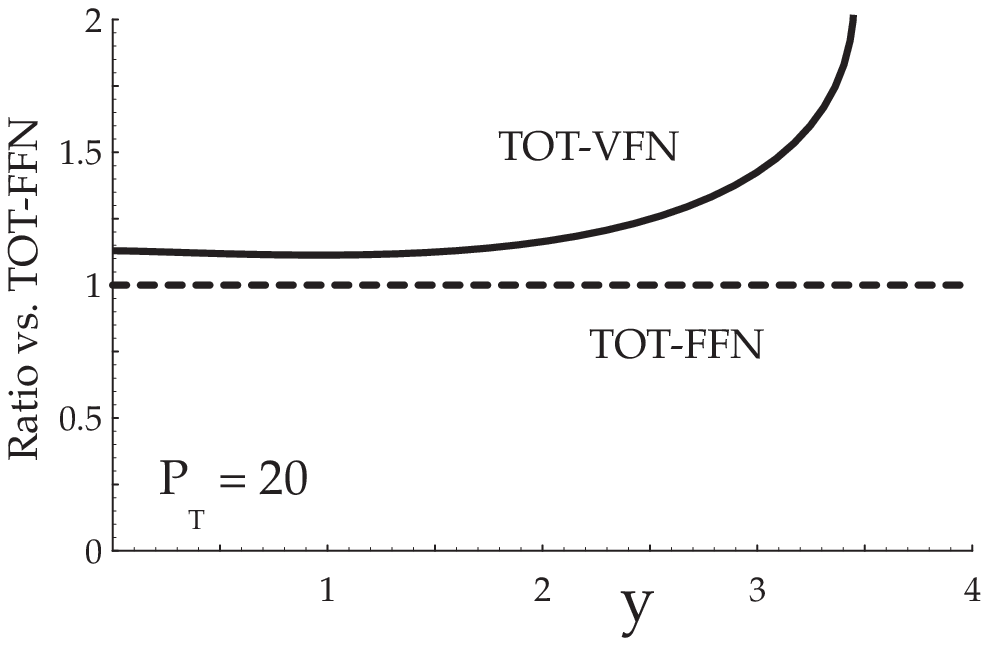}}
 \tightenlines
 \caption{
 Comparison of the total cross section 
 $d\sigma/dp_t^2 /dy$ in the
 ACOT and FFN formalism vs.\ $y$ for $b$ production at 1800 GeV with
 $p_t=20$ GeV and $\mu = M_T$.  
 The curves are scaled by the total FFN cross section
 to facilitate comparison of the relative magnitude.
 }
 \label{fig:Figx9}
\end{figure}
}

\def\figtweax
{
\begin{figure}[hbtp]
 \epsfxsize=0.5\hsize
 \centerline{\epsfbox{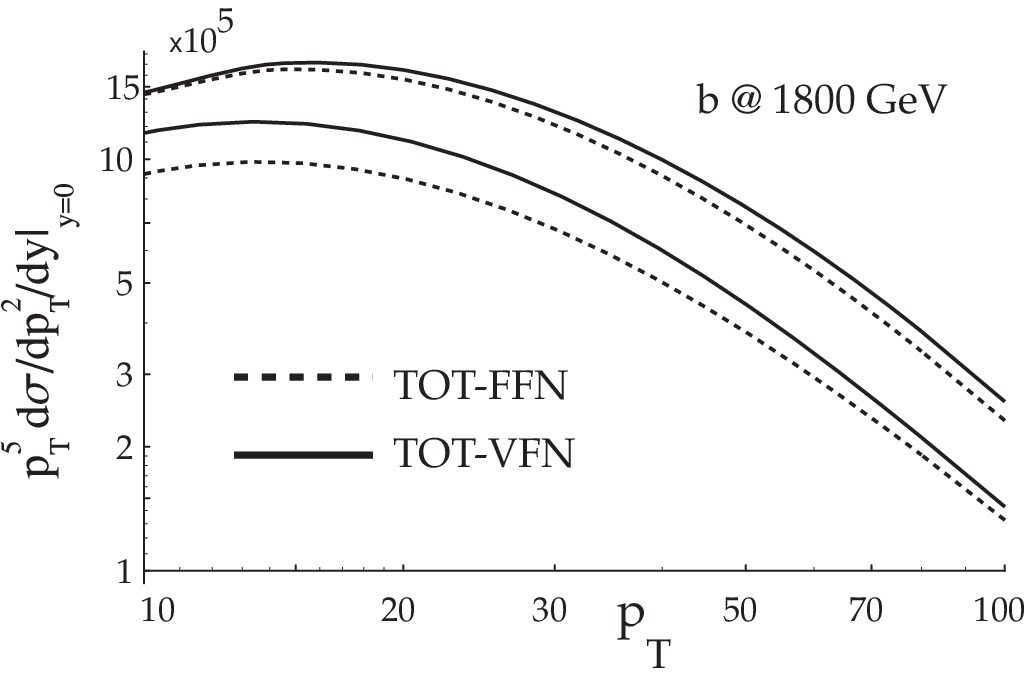}}
 \tightenlines
 \caption{
 Variation of the  total  cross section 
 $p_t^5 \, d\sigma/dp_t^2 /dy|_{y=0}\   (nb\,  {\rm GeV}^{3})$
 in the  ACOT and FFN formalism vs.\ $p_t$ for
 $b$ production at 1800 GeV.
 To gauge the $\mu$ scale variation, we choose
 $\mu=M_T/2$ for upper curves, and
 $\mu=2 M_T$ for lower curves.
 }
 \label{fig:Figx12b1800}
\end{figure}
}

\def\figtwebx
{
\begin{figure}[hbtp]
 \epsfxsize=0.5\hsize
 \centerline{\epsfbox{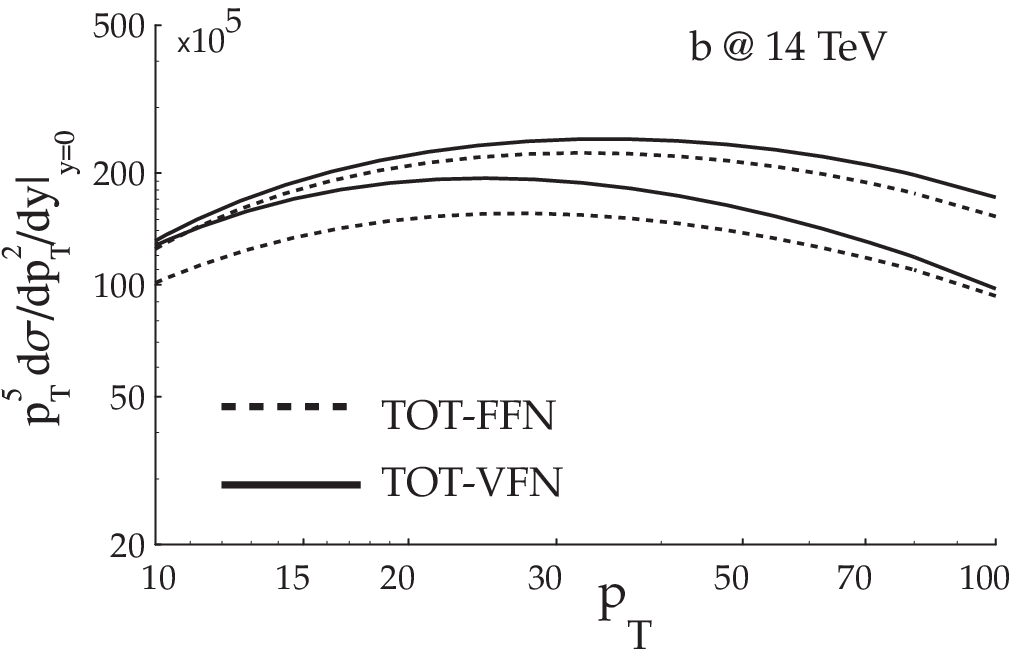}}
 \tightenlines
 \caption{
 Same as Fig.~{\protect \ref{fig:Figx12b1800}}, with $b$ production at 14 TeV.
 }
 \label{fig:Figx12b14tev}
\end{figure}
}

\def\figtwecx
{
\begin{figure}[hbtp]
 \epsfxsize=0.5\hsize
 \centerline{\epsfbox{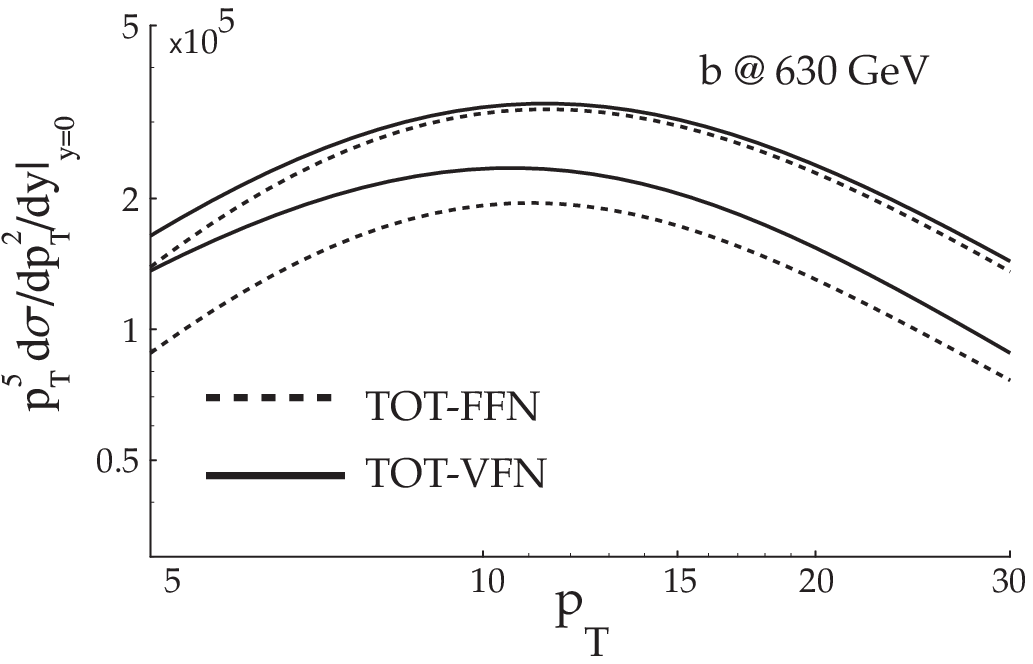}}
 \tightenlines
 \caption{
 Same as Fig.~{\protect \ref{fig:Figx12b1800}}, with $b$ production at 630 GeV.
 }
 \label{fig:Figx12b630}
\end{figure}
}

\def\figxfoteen
{
\begin{figure}[hbtp]
 \begin{center}
 \leavevmode
 \epsfxsize=0.4\hsize \epsfbox{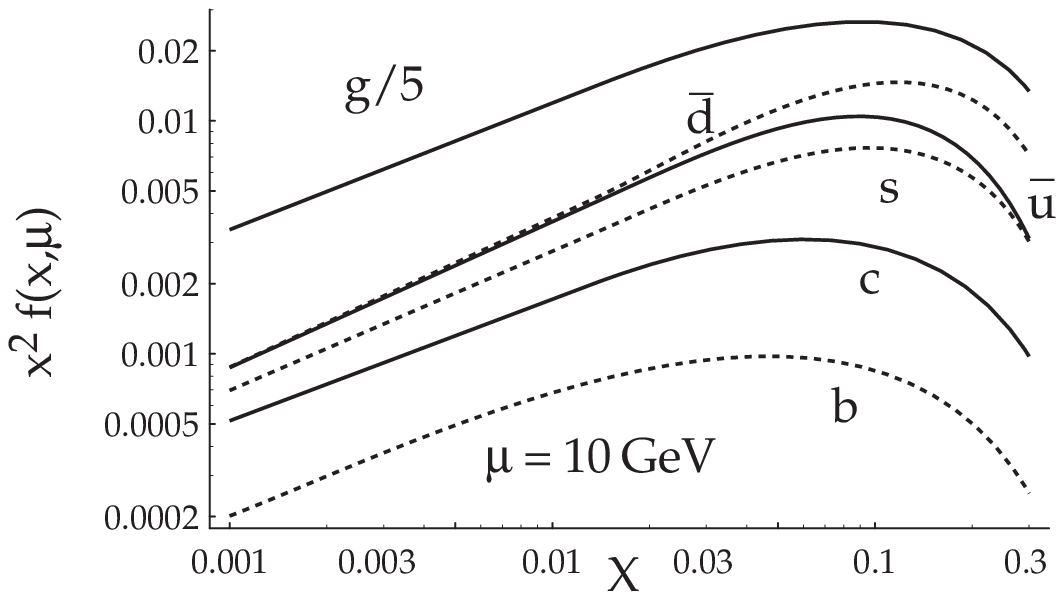} \hfill
 \epsfxsize=0.4\hsize \epsfbox{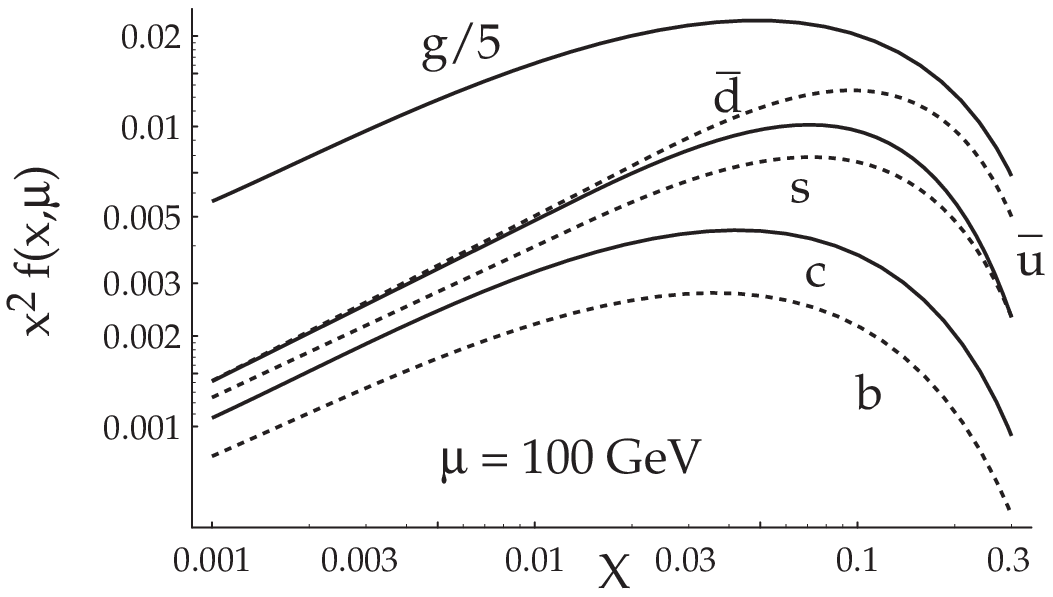}
 \end{center}
 \tightenlines
 \caption{
PDFs vs.\ $x$ for  $\mu$ = 10 and 100 GeV.
The gluon PDF is scaled by 1/5.
 }
 \label{fig:Figx14}
\end{figure}
}

\def\figbexpt
{
\begin{figure}[hbtp]
 \epsfxsize=0.4\hsize
 \centerline{\epsfbox{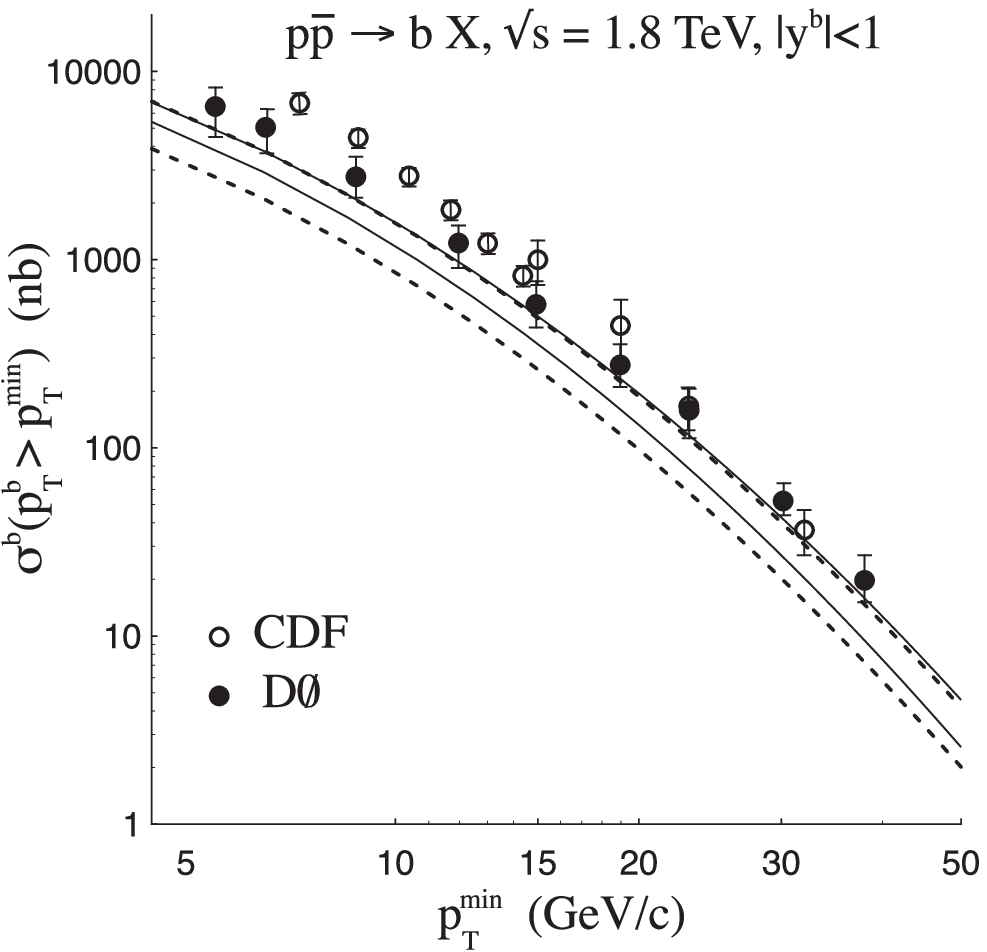}}
 \tightenlines
 \caption{
Comparison with experimental b-production data at the Tevatron 
$\protect\sqrt{s} = 1800$ GeV, $|y|<1$.{\protect \cite{cdf,dzero}}
 The dashed lines represent the $\mu$-variation in the FFN scheme, 
and  solid lines represent the $\mu$-variation in the VFN scheme. 
To gauge the $\mu$ scale variation, we choose
 $\mu=M_T/2$ for upper curves, and $\mu=2 M_T$ for lower curves.
 }
 \label{fig:b_expt}
\end{figure}
}

\begin{document}
\draft
\preprint{hep-ph/9712494}
\title{Heavy Quark Hadroproduction in Perturbative QCD
  \thanks{This work is supported in part by DOE and NSF.}
}
\author{
     F.I. Olness\rlap,$^{a,b}$
%\thanks{Electronic address: olness@mail.physics.smu.edu}
   \ R.J. Scalise\rlap,$^a$
%\thanks{Electronic address: scalise@phys.psu.edu}
   and Wu-Ki Tung$^c$
%\thanks{Electronic address: wkt@cteq06.pa.msu.edu}
}
\address{
   {}$^a$Department of Physics, Southern Methodist Univ., Dallas, TX 75275
\rule{0 em}{3 ex}\newline
   {}$^b$Fermi National Accelerator Laboratory, P.O. Box 500, Batavia, IL 60510
\rule{0 em}{3 ex}\newline
   {}$^c$Department of Physics/Astronomy, Michigan State Univ., East Lansing, 
         MI 48824
\rule{0 em}{3 ex}
}
\date{22 December 1997}
\maketitle
\begin{abstract}
\tightenlines
Existing calculations of heavy quark hadroproduction in perturbative QCD are
either based on the approximate conventional zero-mass perturbative QCD theory
or on next-to-leading order (NLO) fixed-flavor-number (FFN) scheme which is
inadequate at high energies.  We formulate this problem in the general mass
variable-flavor-number scheme which incorporates initial/final state heavy
quark parton distribution/fragmentation functions as well as exact mass
dependence in the hard cross-section.  This formalism has the built-in feature
of reducing to the FFN scheme near threshold, and to the conventional zero-mass
parton picture in the very high energy limit.  Making use of existing
calculations in NLO FFN scheme, we obtain more complete results on bottom
production in the general scheme to order $\alpha_s^3$
both for current accelerator energies and for LHC. The scale dependence of
the cross-section is reduced, and the magnitude is increased with respect
to the NLO FFN results. It is shown that the bulk of
the large NLO FFN contribution to the single heavy-quark inclusive 
cross-section is already contained in the (resummed) order
$\alpha_s^2$ ``heavy flavor excitation'' term in the general scheme.
\end{abstract}

\narrowtext

\section{Introduction}
\label{sec:Intro}

The production of heavy quarks in high energy processes has become an
increasingly important subject of study both theoretically and
experimentally~\cite{FMNR97a}.
The theory of heavy quark production in perturbative Quantum
Chromodynamics (PQCD) is more challenging than that of light parton (jet)
production because of the new physics issues brought about by the additional
heavy quark mass scale. The correct theory must properly take into account the
changing role of the heavy quark over the full kinematic range of the
relevant process from the threshold region (where the quark behaves like a
typical ``heavy particle'') to the asymptotic region (where the same quark
behaves effectively like a {\em parton}, similar to the well known light
quarks $\left\{ u,d,s\right\} $).
Stimulated by significant recent experimental results on heavy quark
production from HERA and the Tevatron, a number of theoretical methods have
been advanced to improve existing QCD calculations of heavy quark production
\cite{ACOT,dis97tung,MRRS,BSV}, incorporating a dynamic role for the
heavy quark parton. The purpose of this paper is to explain how the method
of ACOT \cite{ACOT} is applied to heavy-quark production in hadron-hadron
collisions, to compare results of this approach to existing ``NLO''
calculations, and to demonstrate that it satisfies some important
consistency conditions.

Let us consider the production of a generic heavy quark, denoted by $H$,
with non-zero mass $m_H$, in hadron-hadron collisions. We will define a
quark as ``heavy'' if its mass is sufficiently larger than $\Lambda _{QCD}$
that perturbative QCD is applicable at the scale $m_H$, i.e., that $\alpha
_s(m_H)$ is small. Thus the $c$, $b$, and $t$ quarks are regarded as heavy,
as usual. Let $Q$ be a typical large kinematic variable in the hard
scattering process, in this case the $p_T$ of the heavy quark (or the
associated heavy flavor hadron). The details of the physics of the heavy
quark production process will then depend sensitively on the relative size
of the two scales $m_H$ and $Q$. For simplicity, we will assume there is
only one heavy quark -- $H$ -- that we need to treat. Extending our
treatment to the real world case of $\left\{ c,b,t\right\} $ quarks with
successive higher masses is straightforward.

Conventional PQCD calculations involving heavy quarks consist of two
contrasting approaches: the usual QCD parton formalism uses the {\em %
zero-mass approximation} ($m_H=0$) once the hard scale of the problem (say, 
$Q$) is greater than $m_H,$ and treats $H$ just like the other light partons
\cite{EHLQ,MRS,CTEQ};
on the other hand, most recent ``NLO'' heavy quark production calculations
consider $m_H$ as a large parameter irrespective of the energy scale of the
physical process, and treat $H$ always as a {\em heavy particle}, never as a
parton \cite{CSS85,NDE,Smithetal}. We shall refer to the former as the {\em 
zero-mass variable-flavor-number} (ZM-VFN) scheme -- since the active flavor
number varies, depending on the energy scale $\mu \sim Q$; and the latter as
the {\em fixed-flavor-number} (FFN) scheme -- since the parton flavor number
is kept fixed, independent of $Q$.

Each of these approaches can only be accurate in a limited energy range
appropriate for the approximation involved: $Q\gg m_H$ for the ZM-VFN
scheme; and $Q\sim m_H$ for the FFN scheme. Nonetheless, both approaches
have been used widely beyond their respective regions of natural
applicability: on the one hand, NLO FFN calculations of $c$ and $b$
production are invoked from fixed-target to the highest collider energies
\cite{FMNR97a};
and on the other hand, ZM-VFN calculations dominate practically all other
standard model and new physics calculations, including most global PQCD
analyses (which give rise to commonly used parton distributions \cite
{EHLQ,MRS,CTEQ}) and all
popular Monte-Carlo event generators. The breakdown of the approximations
beyond the original regions of applicability of these approaches can lead to
unreliable results, and introduce large theoretical uncertainties. One
possible sign of the latter is excessive dependence of theoretical
predictions on the unphysical renormalization and factorization scale $\mu$
-- which is well-known to be present in the NLO FFN calculation of
hadro-production cross-section of $c$ and $b$ \cite{SclDep,FMNR97a}.

With steadily improving experimental data on a variety of processes
sensitive to the contribution of heavy quarks (including the direct
measurement of heavy flavor production), it is imperative that the two
diametrically opposite treatments of heavy quarks be reconciled in an
unified framework which can also provide reliable theoretical predictions in
the intermediate energy region, which in reality may well comprise most of
the current experimentally relevant range for charm and bottom physics. This
can be achieved in a general-mass variable-flavor-number (GM-VFN) scheme
which retains the $m_H$ dependence at all energy scales, and which naturally
reduces to the two conventional approaches in their respective region of
validity \cite{ColTun,OlnTun,AOT,ACOT}.
The method is a development of the one devised by Collins, Wilczek, and Zee
\cite{CWZ}. The key point is that one can resum (and factor out) the mass
singularities associated with the heavy quark mass $m_H$ into $H$ parton
distribution and fragmentation functions {\em without simultaneously taking
the } $m_H\rightarrow 0${\em \ limit} in the remaining infra-safe hard
cross-section in the overall physical cross-section formula (as is routinely
done in the ZM-VFN scheme\footnote{%
\tightenlines
The resummation of mass singularities into parton distributions and taking
the zero-mass limit on the hard cross-section are usually performed
simultaneously not as a matter of principle, but for the incidental reason
that the mass singularities are most conveniently identified by using
dimensional regularization in the zero-mass theory.}). The resulting general
formalism represents the natural extension of the familiar PQCD framework to
include {\em heavy quark partons} both in the initial and final states of
the hard scatterings which contribute to high energy processes in the SM and
beyond.

The principles and the practical application of this method were described
in some detail for (heavy quark parton contribution to) Higgs production in
Ref.~\cite{OlnTun} and for lepto-production of heavy quarks in Ref.~\cite
{AOT,ACOT}. It has been applied to the analysis of charm production in
neutrino scattering \cite{CCFRcharm}, and, recently, to a new global QCD
analysis of parton distributions \cite{LaiTun97a}. In the present paper, we
apply this general formalism to heavy quark production in hadron-hadron
colliders. For a concise summary of this formalism, and related recent
developments, see Ref.~\cite{dis97tung}; for a systematic proof of the
factorization theorem which provides the theoretical foundation of this
formalism, see Ref.~\cite{Collins97}. In recent literature on
leptoproduction of heavy quarks, there have been two other formulations of
the VFN scheme with non-zero heavy quark mass: an order $\alpha_s$ scheme by
Ref.~\cite{MRRS} and an order $\alpha_s^2$ scheme by Ref.~\cite{BSV}. For
definiteness, we shall refer to our implementation of the general principles
as the ACOT scheme. Although not unique, it represents in many ways the
simplest and the most natural one in relation to the familiar
\mbox{\small
{$\overline {\rm MS}$}\ } ZM-VFN scheme \cite{dis97tung}. See Ref.~\cite
{dis97oln} for comments on some aspects of scheme dependence (in particular,
mass-dependent or mass-independent evolution of the parton distributions),
as well as on comparison to the scheme proposed by Ref.~\cite{MRRS}.

The following section presents details of our calculation scheme as applied
to hadroproduction and describes the physical origin of the various terms
which appear in the formalism. Sec.~\ref{sec:Ingredients} contains specific
information on how the various ingredients of the formalism are calculated.
This is followed by numerical comparisons of the new calculation with existing
FFN scheme results on the inclusive differential $p_T$ cross
section for bottom production at current accelerator and
LHC energies. In the concluding section, we discuss what remains to be done
for a full understanding of heavy quark production in PQCD.

\section{Hadro-production of $H$ in the general mass formalism}

\label{sec:formalism}

Consider the hadro-production of a generic heavy quark $H$:
\begin{equation}
A+B\longrightarrow H+X  \label{process}
\end{equation}
where $A$ and $B$ are hadrons, and $X$ includes $\overline{H}$ along with
all other summed-over final state particles. In the ACOT formalism for
heavy quark production developed in Refs.~\cite{ColTun,ACOT,Collins97}, the
inclusive cross section for observing $H$ with a given momentum $p$ at
high energies is given by a factorization formula of the same form
as in the familiar zero-mass QCD parton formalism:
\begin{equation}
\sigma _{AB}^{HX}=\sum_{a,b,c}f_{AB}^{ab}\otimes \hat{\sigma}%
_{ab}^{c,r}\otimes d_c^H ,  \label{FacThmH}
\end{equation}
where $\left\{ A,B\right\} $ denote the initial state hadrons, $\left\{
a,b\right\} $ the initial state partons, $f_{AB}^{ab}$ the associated parton
distribution functions in the combination $f_{AB}^{ab}\equiv f_A^af_B^b$, 
$\hat{\sigma}_{ab}^{c,r}$ the perturbatively calculable {\em infra-red safe}
hard-scattering cross section for $\left\{ a,b\right\} \rightarrow \left\{
c,r\right\} $ which is free of large logarithms of $m_H$ over the full
energy range, and $d_c^H$ the fragmentation functions for finding $H$ in $c$. 
All {\em active partons} are included in the summation over $\left\{
a,b\right\} $ and $\left\{ c\right\} $, including $H$ provided the
renormalization and factorization scale $\mu $ is larger than $m_H$.%
\footnote{%
\tightenlines
For simplicity, we shall use the symbol $\mu $ to represent collectively the
renormalization scale as well as the factorization scales for parton
distributions and for fragmentation functions.}

Since we shall compare our results to those of the FFN scheme, it is
necessary to draw the distinction between the heavy quark $H$ and the
associated {\em light partons}, we shall denote the latter collectively by 
$l$ for simplicity of notation. By definition, $l=\left\{ g,q\right\} $
where $g$ is the gluon and $q$ denotes the light quarks in the sense of the
FFN scheme: $q=\left\{ u,d,s\right\} /\left\{ u,d,s,c\right\} $ for
charm/bottom production respectively. The number of light quark partons will
be denoted by $n_l.$

As mentioned in the introduction, an important distinguishing feature of
this general formalism from the familiar ZM-VFN parton approach is that,
after subtraction of mass-singularities, the full $m_H$ dependence in the
perturbatively calculated hard scattering coefficients $\hat{\sigma}%
_{ab}^{c,r}$ is retained. This allows the theory to maintain accuracy and
reproduce the FFN scheme results in the threshold region, as required by
physical considerations. On the other hand, our parton densities (and
fragmentation functions) are defined in the
\mbox{\small {$\overline {\rm
MS}$}\ }{} scheme, hence they satisfy mass-independent evolution equations
--- the same as in the usual zero-mass formalism. Considerable simplification
then results in the implementation of this scheme since the well-established
NLO evolution kernels and evolution programs can be directly used. It should
be noted, however, that the parton densities do have implicit dependence on
quark masses: to the leading power in $\Lambda /m_H$, this dependence is
generated by the matching conditions at $\mu =m_H$ between the parton
densities below and above the threshold\footnote{\label{fn:radiativeChm}
\tightenlines
Note that equivalent matching conditions can be derived for any $\mu \ $%
which is of order $m_H$. Also, possible non-perturbative heavy quarks 
\cite{intrinsic}, as opposed to ``radiatively generated'' ones (assumed in
subsequent discussions), can be incorporated in the general scheme by
allowing for a nonzero heavy quark density in the below-threshold part of
the scheme.} --- they are defined in each of the two regions by the
respective renormalization scheme adopted for that region 
\cite{ColTun,Collins97}.

The first few terms in the perturbative expansion of the production cross
section, Eq.~\ref{FacThmH}, are schematically
\begin{equation}
\begin{array}{rllrr}
\sigma _{AB}^{HX} & = & f_{AB}^{ll^{\prime }}\,\otimes \,^2\hat{\sigma}%
_{ll^{\prime }}^{H\bar{H}}\otimes \,d_H^H & : & \text{HC0} \\
&& +  f_{AB}^{lH}\,\otimes \,^2\hat{\sigma}_{lH}^{lH}\,\otimes d_H^H
   +  f_{AB}^{Hl}\,\otimes \,^2\hat{\sigma}_{Hl}^{lH}\,\otimes d_H^H
& : &
\text{HE1} \\
&& +  \,f_{AB}^{ll^{\prime }}\,\otimes \,^3\hat{\sigma}_{ll^{\prime }}^{H%
\bar{H}g}\otimes \,d_H^H & : & \text{HC1} \\
&& +  f_{AB}^{ll^{\prime }}\,\otimes \,^2\hat{\sigma}_{ll^{\prime
}}^{l_1l_2}\otimes d_{l_1}^H+f_{AB}^{Hl}\,\otimes \,^2\hat{\sigma}%
_{Hl}^{Hl}\otimes d_l^H & : & \text{GF1,2} \\
&& +  f_{AB}^{HH}\,\otimes \,^2\hat{\sigma}_{HH}^{HH}\otimes d_H^H & : &
\text{HH2}
\end{array}
\label{HadXsec1}
\end{equation}
where repeated indices $l$, $l^{\prime }$, $l_1$, and $l_2$ are summed over
all light flavors ($q$ and $g$) and the pre-superscript $n$ on $^n\hat{%
\sigma}$ denotes the formal order in the expansion of $\alpha _s$.

The physical interpretation of the various terms, labeled by the
abbreviations in the last column of this equation, can be made apparent by
the generic diagrams shown in Fig.~\ref{fig:HadXsecA}. They are:

\begin{itemize}
\item[(i)]  HC0: the order $\alpha _s^2$ {\em heavy flavor creation} process
$l~l^{\prime }\rightarrow H\,\bar{H}$ followed by heavy-quark fragmentation;

\item[(ii)]  HE1: the order $\alpha _s^2$ {\em heavy flavor excitation}
process $l~H\rightarrow l~H$ followed by heavy-quark fragmentation;

\item[(iii)]  HC1: order $\alpha _s^3$ virtual and real corrections to HC0, 
$l~l^{\prime }\rightarrow H\bar{H}g,$ with fragmentation;

\item[(iv)]  GF1,2: order $\alpha _s^2$ (light- and heavy-) parton-gluon
scattering $l~g\rightarrow l~\,g$ and $H~g\rightarrow H~g,$ followed by 
{\em gluon fragmentation}\footnote{%
\tightenlines
For generality, we have written the summation (over $l$) to include all
light-parton fragmentation into $H$. In fact, we expect gluon fragmentation
to dominate. Light-quark fragmentation is suppressed because of the quark
structure of the evolution equation. In the fixed order calculations which
we will do to compare schemes, fragmentation of light quarks $q$ and heavy
antiquarks $\bar{H}$ into $H$ are both at least of order $\alpha _s^2$.}
into the heavy quark $H$; and

\item[(v)]  HH2: the order $\alpha _s^2$ $HH$ {\em scattering} $HH(\bar{H}%
)\rightarrow HH(\bar{H}),$ with fragmentation.
\end{itemize}

\figHadXsecA
For the diagrams in Fig.~\ref{fig:HadXsecA} we use a dashed line for all
light partons $l$ (representing collectively gluons and light quarks) to
distinguish them from the heavy quark $H$, represented by a solid line. The
square blocks in these diagrams represent the hard cross sections $^n\hat{%
\sigma}$, obtained from all Feynman diagrams of the same order with the
external lines indicated. The round blobs represent parton distribution and
fragmentation functions. The initial state hadron lines for the parton
distribution factors are suppressed in all the diagrams. For the
differential cross section $d\sigma /dp_T^H,$ the four terms involving $%
2\rightarrow 2$ hard scattering processes are tree-level processes. Beyond
tree level, we have included only the most important order $\alpha _s^3$
term --- HC1 --- which contains $2\rightarrow 3$ real and $2\rightarrow 2$
virtual corrections to the\ leading HC0 cross-section. This term corresponds
to the next-to-leading (NLO) contribution in the FFN scheme; it is known to
be large and it is the only one that has been calculated to this order so
far (see below). We will discuss the relative sizes of the various terms in
the next paragraph, and comment on the possible significance of other terms
not included here in later parts of the paper.

In conventional applications of PQCD with light partons, it is common to
distinguish the order of terms in a cross section by treating the
(non-perturbative) parton distributions in $f_{AB}^{ab}$ as all being of
order one, and then counting powers of $\alpha _s$ in the hard scattering.
With the inclusion of heavy quark partons, one may expect heavy quark
distribution and fragmentation functions to be of order $\alpha _s$ in
the energy region not too far above threshold,
assuming they are purely radiatively generated.\footnote{%
\tightenlines
As mentioned in footnote \ref{fn:radiativeChm}, our formalism can
accommodate non-perturbative, i.e.\ non-radiative or ``intrinsic" heavy
quarks. However, we shall not get into that possibility in this paper.} We
know that at large $x$ the valence $u$ and $d$ quarks dominate all the other
partons; while at small $x,$ it is the gluon density that dominates. For
instance, in the latter region, we compare in Fig.~\ref{fig:Figx14} the
relative sizes of the various sea partons at two energy scales to the gluon
(scaled down by a factor of 5). In all cases it appears safe to regard the
fully evolved heavy quark density $f_A^H$ to be of effective order $\alpha
_s $ with respect to the dominant parton density (gluon or valence quark),
i.e.,
\begin{equation}
f_A^H\sim {\cal O}(\alpha _s)  \label{orderf}
\end{equation}
Note, however, that the evolved charm density comes within a factor of 2 of
the other sea quark densities at small $x$.

\figxfoteen

The same argument leads to the following expectations for the fragmentation
functions (note, we confine ourselves to the production of the heavy flavor
parton $H,$ rather than the associated hadron):
\begin{equation}
\begin{array}{lll}
\,d_H^H & \sim & \delta (1-z)+{\cal O}(\alpha _s) \\
d_g^H & \sim & {\cal O}(\alpha _s) \\
d_{q,\bar{H}}^H & \sim & {\cal O}(\alpha _s^2)
\end{array}
\label{orderd}
\end{equation}
These order-of-magnitude estimates are, of course, expected to become
inapplicable at very large scales. The gluon-to-heavy-quark fragmentation
function, in particular, can become substantial at large scales.

Nevertheless, let us use these estimates as the first guide to orders of
magnitudes of the terms in Eq.~\ref{HadXsec1}. The first term (HC0) is of
order $\alpha _s^2$; the HE1 and GF1 terms, and the HC1 term are of
effective order $\alpha _s^3$; and the GF2 and the HH2 terms are of
effective order $\alpha _s^4$. (This naive counting of effective order
explains the choice of numerical suffixes in the labels for these terms.)
The actual relative numerical importance of the various terms will also
depend on other considerations such as large color factors or dynamical
effects (e.g.\ spin-1 $t$-channel exchange contribution at high energies).
This is well-known for the perturbative expansion of various processes,
including heavy quark production calculated in the FFN scheme, where the
order $\alpha _s^3$ NLO term (corresponding to HC1 without heavy quark mass
subtractions) is actually larger than the order $\alpha _s^2$ LO term (HC0).
We will examine in detail the numerical significance of the various terms in
Sec.~\ref{sec:Results}. The results are revealing, since they show simple
features which are not accessible from the conventional FFN scheme
perspective.

To specify fully the calculation in our scheme, we need to specify the
perturbative hard cross sections $\hat{\sigma}_{ab}^{c,r}$. Following Ref.~%
\cite{ACOT}, we obtain these by applying the factorization formula, Eq.~\ref
{FacThmH}, to cross-sections involving partonic beams, that is to $\sigma
_{ab}^{c,r}$ (without the caret) calculated from the relevant Feynman
diagrams. Then we solve for the hard cross sections $\hat{\sigma}_{ab}^{c,r}$
order by order in $\alpha _s$. Specifically, in a given order $\alpha _s^n,$
the factorization theorem implies that the partonic cross section has the
form
\begin{equation}
^n\sigma _{\alpha \beta }^{HX}=\sum \;^{n_1}f_{\alpha \beta }^{ab}\otimes
\,^{n_2}\hat{\sigma}_{ab}^{c,r}\otimes \,^{n_3}d_c^H ,  \label{FacThmP}
\end{equation}
where the sum is over the values of $n_1$, $n_2$, and $n_3$ that give the
correct order, i.e., $n=n_1+n_2+n_3$. On the right-hand side, this equation
differs from Eq.~\ref{FacThmH}
in two respects.  First, the parton distributions are relative to
an on-shell parton target, instead of a hadron target.  Secondly,
we have expanded the parton-level distribution and
fragmentation functions in powers of $\alpha_s$, with, for example,
%% JCC
$^{n_1}f_\alpha ^a$ denoting the term of $\alpha_s^{n_1}$
in the distribution function.  These coefficients
are calculated order by order in $\alpha _s$, but are not
infra-red safe; they are used only in the intermediate steps toward the
derivation of $\hat{\sigma}_{ab}^{c,r}$.

The most general method to obtain the hard scattering coefficients is
directly from Eq.\ \ref{FacThmP}.
%% JCC
All of the full partonic cross sections, the
left-hand side, and the parton densities and fragmentation functions at the
partonic level, on the right-hand side, can be computed from definite
%%JCC
Feynman rules.  From them, Eq.\ \ref{FacThmP} gives the hard scattering
coefficients. However, at the order $\alpha_s^3$ level to which we are
working, it is convenient to carry out this calculation in two steps; the
relation to the coefficient functions of the other schemes in the
literature is then readily obtained.

First, the unsubtracted cross-sections are calculated from all relevant
Feynman diagrams with the use of dimensional regularization (with the
familiar parameters $\epsilon $ and $\mu $), with all light quark masses set
to zero, and with the heavy quark mass non-zero. After ultra-violet
renormalization (with \mbox{\small $\overline {\rm MS}$} counter terms) and
cancellation of infra-red divergences (after combining real and virtual
diagrams), the cross-section formula will depend on the renormalized mass 
$m_H$ and the unphysical parameters ($\epsilon$ and $\mu$), in addition to
the kinematic variables. The collinear singularities associated with the
$1/\epsilon$ poles can be factorized into a kind of {\em distribution
function of light partons in a light parton} $\bar{f}_\alpha
^a(\epsilon ,\mu )$ using the \mbox{\small {$\overline {\rm MS}$}\ }
convention. Since the cross-section is inclusive with respect to the light
partons, we need no light parton fragmentation functions. Thus,\footnote{%
\tightenlines
For simplicity, we have used the symbol $\mu$ to collectively represent
both the factorization scale and the renormalization scale.}
\begin{equation}
^n\sigma _{\alpha \beta }^{HX}(m_H,\epsilon ,\mu )=\sum_{n_1+n_2=n}\;^{n_1}%
\bar{f}_{\alpha \beta }^{ab}(\epsilon ,\mu )\otimes \,^{n_2}\tilde{\sigma}%
_{ab}^{HX}(m_H,\mu ),  \label{FacThmP1}
\end{equation}
where $\bar{f}_{\alpha \beta }^{ab}$ represents the product of two singular
light
parton distributions (cf.\ Eq.~\ref{FacThmH}), and all non-essential
(kinematical and convolution) variables have been suppressed. This concludes
the first step of the construction, where the set of {\em intermediate
finite cross-sections} $^n\tilde{\sigma}_{ab}^{HX}(m_H,\mu
)$ are obtained by systematically subtracting the collinear singularities.

At first sight, this appears to imply that the sub-set of $^n\tilde{\sigma}%
_{ab}^{HX}(m_H,\mu )$ with $\left\{ a,b\right\} =\left\{ l,l^{\prime
}\right\} $ (light partons) are the same as the conventional FFN scheme
cross-sections. This is in fact true at the lowest non-trivial order, which
is all that we will consider in this paper. However, at sufficiently high
order extra heavy-quark loops come in,
and these are renormalized differently in the
FFN scheme and in the $\overline{{\rm MS}}$ scheme that we use when $\mu >
m_H$. Hence the singularities to be subtracted differ by some kind of
renormalization-group transformation. In this paper, we do not treat these
higher order graphs.

The second stage of our calculation starts from the observation that
although they are finite, the cross-sections $^n\tilde{\sigma}%
_{ab}^{HX}(m_H,\mu )$ contain logarithmic mass-singularities, i.e.\ powers of
%% JCC
$\ln \left( m_H/\mu \right)$, in the $m_H/\mu \rightarrow 0$
limit. The second step of the derivation consists of factoring these
singularities out to arrive at the {\em fully infra-red safe hard
cross-sections} $^n\hat{\sigma}_{ab}^{c,r} $ of Eq.~\ref{FacThmP}.
Explicitly,
\begin{equation}
^n\tilde{\sigma}_{\alpha \beta }^{HX}(m_H,\mu )=\sum \;^{n_1}\tilde{f}%
_{\alpha \beta }^{ab} \left(\ln (m_H/\mu) \right) \otimes \,^{n_2}\hat{\sigma%
}_{ab}^{c,r}(m_H,\mu )\otimes \,^{n_3} \tilde d_c^H \left(\ln ( m_H/\mu
)\right) .  \label{FacThmP2}
\end{equation}
Here the logarithmically singular terms in the $%
m_H\rightarrow 0$ limit are factored into $\tilde{f}$ and $\tilde d$. These
are partonic level parton densities and fragmentation functions with
subtractions made to remove the singularities associated with light partons.
These subtractions exactly correspond to the subtractions used to obtain 
$\tilde\sigma$ from $\sigma$. The remaining infra-red safe $m_H$ dependence
is kept in $\hat{\sigma}$ (in contrast to the conventional approach, where 
$m_H$ is set to zero).

As with all calculations of hard scattering coefficients, there is a freedom
to choose exactly how to define the parton densities and fragmentation
functions. The choice defines the factorization scheme, cf.\ \cite{dis97oln}%
, and thus determines how much of the finite $m_H$ dependence is included in
$\tilde{f}$ and $\tilde d.$

We use the ACOT scheme \cite{ColTun,ACOT,Collins97}. In this scheme, the
parton densities and fragmentation functions in the region $\mu>m_H$ are
determined by the requirement that all ultra-violet divergences are
renormalized by the $\overline{{\rm MS}}$ scheme --- both the ultra-violet
divergences in the renormalization of the interactions of QCD and the
ultra-violet divergences that make finite the parton densities (in hadrons).
The hard scattering coefficients $\hat\sigma$ in Eq.\ \ref{FacThmP} are then
well-defined. In the first stage of our calculation, the factorization of
light parton singularities, we define the intermediate coefficients $%
\tilde\sigma$ in Eq.\ \ref{FacThmP1} by subtraction of collinear
singularities in the $\overline{{\rm MS}}$ scheme.

Subtracting the terms which contain $\tilde{f}$ and $\tilde d$ with
non-trivial mass singularities from $^n\tilde{\sigma}_{\alpha
\beta}^{HX}(m_H,\mu )$, we obtain the {\em fully infra-red safe hard
cross-sections} $^n\hat{\sigma}_{ab}^{c,r}(m_H,\mu )$ which we need in Eq.~%
\ref{FacThmH}. The relevant non-vanishing perturbative parton distributions 
$^n\tilde{f}$ up to order $\alpha _s,$ are:
\begin{equation}
\begin{array}{rll}
^0\tilde{f}_a^b(x) & = & \delta _a^b\,\delta (1-x) \\
{}^1\tilde{f}_a^H(x,\mu ) & = & \frac{\alpha _s(\mu )}{2\pi }\ \ln \left(
\frac{\mu ^2}{m_H^2}\right) \ P_{a\to H}(x) \\
^1\tilde{f}_g^g(x,\mu ) & = & \frac{\alpha _s(\mu )}{2\pi }\ \ln \left(
\frac{\mu ^2}{m_H^2}\right) \ \delta (1-x)
\end{array}
\label{pertpdf}
\end{equation}
where $a,b=\{g,H\}$, $P_{a\to b}(x)$ is the usual first order splitting
function. The only nonzero terms at order $\alpha_s$ in Eq.\ \ref{pertpdf}
come from contributions with a heavy quark loop. All one-loop corrections
that involve light partons are zero, because of the well-known cancellation
%%JCC
of IR and UV singularities. For the fragmentation functions, we have
\cite{MelNas91}:
\begin{equation}
\begin{array}{rll}
   {}^0\tilde{d}_H^H(z) & = & \delta (1-z)
\\
   {}^1\tilde{d}_H^H(z) & = &
    \frac {\alpha _s C_F}{2\pi }
    \left[
        \frac {1+x^2}{1-x}
        \left(
            \ln \frac {\mu ^{2}}{M_H^2} - 2 \ln (1-x) -1
        \right)
     \right]_+
\\
   {}^1\tilde{d}_g^H(z,\mu ) & = &
          \frac{\alpha _s(\mu )}{2\pi }\
          \ln \left( \frac{\mu ^2}{m_H^2}\right) \
          P_{a\to b}(z)
\end{array}
\label{pertfrg}
\end{equation}

We can now substitute Eqs.\ \ref{pertpdf} and \ref{pertfrg} into Eq.~\ref
{FacThmP2} to solve for $^n\hat{\sigma}_{ab}^{c,r}.$ To order $\alpha _s^2$,
only $^0\tilde{f}_a^b$ and $^0\tilde{d}_a^b$ are needed and we obtain the
trivial identities,
\begin{equation}
\begin{array}{lll}
^2\tilde{\sigma}_{l\,l^{\prime }}^{H\bar{H}} & = & ^2\hat{\sigma}%
_{l\,l^{\prime }}^{H\bar{H}} \\
^2\tilde{\sigma}_{lH}^{lH} & = & ^2\hat{\sigma}_{lH}^{lH} \\
^2\tilde{\sigma}_{HH}^{HH} & = & ^2\hat{\sigma}_{HH}^{HH} \\
^2\tilde{\sigma}_{H\bar{H}}^{H\bar{H}} & = & ^2\hat{\sigma}_{H\bar{H}}^{H%
\bar{H}}
\end{array}
\label{loPxsec}
\end{equation}
which reflect the fact that the order $\alpha _s^2$ cross sections are given
%% JCC
by tree graphs: i.e.\ the $^2\tilde{\sigma}$'s are infra-red safe,
in the $m_H \to 0$ limit.  Hence they
do not need any subtraction. Following the same procedure to order $\alpha
_s^3,$ we obtain schematically:\footnote{%
\tightenlines
It should be observed that the only contribution to ${}^1\tilde{f}_g^g$, in
Eq.\ \ref{pertpdf}, corresponds to a gluon self-energy graph on the external
gluon line. Therefore when obtaining $\hat\sigma$ from $\tilde\sigma$ in
Eq.\ \ref{nloPxsec}, the effect of the ${}^1\tilde{f}_g^g$ terms is simply
to cancel external line corrections.}
\begin{eqnarray}
{}{}\Sigma _X\,^3\tilde{\sigma}_{l\,l^{\prime }({\rm real+virtual})}^{H\bar{H%
}X} &=&{}\Sigma _X\,{}^3\hat{\sigma}_{l\,l^{\prime }}^{H\bar{H}X}  \nonumber
\\
&&+\left[ {}^1\tilde{f}_g^g\,\otimes \,{}^2\hat{\sigma}_{gl^{\prime }}^{H%
\bar{H}}+{}^1\tilde{f}_g^g\,\otimes \,{}^2\hat{\sigma}_{lg}^{H\bar{H}%
}\right] +{}^2\hat{\sigma}_{l\,l^{\prime }}^{H\bar{H}}\otimes \,^1\tilde{d}%
_H^H  \nonumber \\
&&+\left[ {}^1\tilde{f}_{l^{\prime }}^H\otimes {}^2\hat{\sigma}%
_{lH}^{lH}+{}^1\tilde{f}_l^H\otimes {}^2\hat{\sigma}_{Hl^{\prime
}}^{Hl^{\prime }}\right] +{}^2\hat{\sigma}_{ll^{\prime }}^{l_1l_2}\,\otimes
\,^1\tilde{d}_{l_1}^H  \label{nloPxsec}
\end{eqnarray}
For convenience, we have used a single equation to cover the various
possible initial states. Not all terms on the right-hand side are applicable
to all cases: for gluon-gluon scattering, $\left\{ l\,l^{\prime }\right\}
=\left\{ gg\right\} $, all terms are present; for $\left\{ l\,l^{\prime
}\right\} =\left\{ q\bar{q}\right\} ,$ only the 1st, 3rd and 5th terms
contribute; and for $\left\{ l\,l^{\prime }\right\} =\left\{ gq(\bar{q}%
)\right\} ,$ only the 1st, 4th and 5th terms contribute.

This equation can easily be inverted to obtain the order $\alpha _s^3$ hard
cross section in terms of the calculated finite intermediate cross-sections 
$^3\tilde{\sigma}$ (with non-zero mass $m_H)$ and various subtraction terms
which remove the mass singularities associated with the heavy quark degree
of freedom:
\begin{equation}
\begin{array}{lllrr}
{}\Sigma _X\,^3\hat{\sigma}_{l\,l^{\prime }}^{H\bar{H}X} & = & {}\Sigma
_X\,{}^3\tilde{\sigma}_{l\,l^{\prime }({\rm real+virtual})}^{H\bar{H}X} &
\hspace{0.16in}: & \mbox{HC1-FFN} \\
&  & -{}^1\tilde{f}_g^g\otimes {}^2\tilde{\sigma}_{gg^{\prime }}^{H\bar{H}%
}-{}^1\tilde{f}_{g^{\prime }}^{g^{\prime }}\otimes {}^2\tilde{\sigma}%
_{gg^{\prime }}^{H\bar{H}} & : & \mbox{Fln-Sub}{} \\
&  & -{}^2\tilde{\sigma}_{l\,l^{\prime }}^{H\bar{H}}\otimes \,^1\tilde{d}_H^H
& : & \mbox{HF1-Sub} \\
&  & -{}^1\tilde{f}_{l^{\prime }}^H\otimes 
{}^2\tilde{\sigma}_{lH}^{lH}-{}^1\tilde{f}%
_{l}^H\otimes {}^2\tilde{\sigma}_{Hl^{\prime }}^{Hl^{\prime }} & :
& \mbox{HE1-Sub} \\
&  & -{}^2\tilde{\sigma}_{ll^{\prime }}^{l_1l_2}\otimes {}^1\tilde{d}_{l_1}^H
& : & \mbox{GF1-Sub}
\end{array}
\label{nloIRSxsec}
\end{equation}
Here, we have replaced all tree-level $^2\hat{\sigma}$ in Eq.~\ref{loPxsec}
by the corresponding $^2\tilde{\sigma}$\ because they are the same, cf.\ Eq.~%
\ref{loPxsec}. The content of the terms on the right-hand-side of this
equation, labeled by the abbreviations in the last column, can be seen more
easily from the diagrams in Fig.~\ref{fig:NloXsec} where, for clarity, the
four possible initial/final state channels are separately shown.

\figNloXsec

The terms in Eq.~\ref{nloIRSxsec} and Fig.~\ref{fig:NloXsec} are:

\begin{itemize}
\item  HC1-FFN : the usual order $\alpha _s^3$ FFN scheme result, due to
contributions of the NLO-HC (virtual and real) diagrams, with infra-red and
collinear singularities associated with light partons cancelled/subtracted
in the conventional way (in the \mbox{\small {$\overline {\rm MS}$}\ }
scheme);

\item  Fln-SUB : the correction to the order $\alpha _s^2$ gluon-gluon HC
process due to the difference in the definition of the gluon distribution
between the below-threshold ($n_f=n_l$) scheme and the above-threshold 
($n_f=n_l+1$) scheme;

%% JCC
\item  HE-SUB : the subtraction of large logarithms of the heavy mass
contained in $^1\tilde{f}_g^H,$ cf.~\ Eq.~%
\ref{pertpdf}, due to a flavor-excitation configuration;

%% JCC
\item  HF-SUB (GF-SUB) : the subtraction of large logarithms of the
heavy mass in the final state heavy-quark (gluon) fragmentation
residing in $^1\tilde{d}_{H,g}^H,$ cf.\ Eq.~\ref{pertfrg}.
\end{itemize}

After these subtractions, $^3\hat{\sigma}_{l\,l^{\prime }}^{H\bar{H}g}$ is
free from all large logarithms associated with potential mass singularities,
%% JCC
i.e.\ it is infra-red safe with respect to $m_HQ\rightarrow 0$. When
this result is used in Eq.~\ref{HadXsec1}, the hadronic cross section is
well-behaved in the high energy limit, in contrast to the FFN scheme result
which would diverge because of the large logarithms. In fact, in this limit
Eq.~\ref{HadXsec1} reduces to the usual zero-mass $\overline{{\rm MS}}$ $%
\alpha _s^3$ parton formula with the $H$ quark counted just like the other
light partons. This is, of course, the correct limit at high energies.

Some insight can be gained by explicitly substituting Eqs.~\ref{loPxsec},
\ref{nloIRSxsec} in Eq.~\ref{HadXsec1} and obtaining the hadronic cross
section in terms of the intermediate partonic cross-sections $\left\{ ^2%
\tilde{\sigma},^3\tilde{\sigma}\right\} $ (which are all finite for non-zero
$m_H$) along with the necessary subtraction terms which represent the
overlap between the two sets of cross-sections and which remove the
potentially dangerous mass singularities. The full set of terms are most
clearly displayed in diagrammatic form, as shown in Fig.~\ref{fig:HadXsecB}.
\figHadXsecB
The first column lists contributions from all the $2\rightarrow 2$
tree-level diagrams summarized in Eq.~\ref{HadXsec1} and Fig.~\ref
{fig:HadXsecA} plus 1-loop virtual corrections from Eq.~\ref{nloIRSxsec};
the last column lists contributions from all the $2\rightarrow 3$ terms
contained in Eq.~\ref{nloIRSxsec}; and the middle column contains all the
relevant subtraction terms. These terms can most easily be obtained by
substituting the terms shown in Fig.~\ref{fig:NloXsec} in those in Fig.~\ref
{fig:HadXsecA}, summing $l$ over $\left\{ g,q\right\} $. With the exception
of uniformly suppressing the initial state parton distribution factors
(represented by dark blobs in Fig.~\ref{fig:HadXsecA}), these diagrams
contain all the ingredients needed to write down the full formula for the
cross-section. In between the last two columns, we have also drawn lines to
indicate the origins of the various subtraction terms in the $2\rightarrow 3$
diagrams, according to Fig.~\ref{fig:NloXsec}.

In this way of organizing the results, the order $\alpha _s$ subtraction
terms in Fig.~\ref{fig:HadXsecB} are shown next to the relevant HE/GF/HF
contributions in the same row, making explicit the physical origin of the
subtractions: these terms containing large-logarithm (present in the
un-regulated FFN scheme calculations) also represent the low-order
components of the QCD evolved parton distribution/fragmentation functions.
As an example, consider the first two terms in row one of HE1:
\begin{equation}
{\rm (2}\rightarrow {\rm 2}\tilde{\sigma})-(m_H{\rm -subtraction}){\Big |}_{%
{\rm HE1}}^{gH\rightarrow gH}=f_A^g(f_A^H-f_A^g\,\otimes \,^1\tilde{f}%
_g^H)\otimes \,^2\tilde{\sigma}_{gg}^{H\bar{H}}\otimes d_H^H
\label{MassSubtr}
\end{equation}
Both from this formula and from the corresponding graphs, one can see that
the subtraction term containing $f_A^g\,\otimes \,^1\tilde{f}_g^H$
represents that part of the NLO-FFN contribution which is already included
in the (fully evolved) parton distribution $f_A^H$. The latter, of course,
represents the result of resumming all powers of $\alpha _s\ln \left( \mu
^2/m_H^2\right) $, hence contains important physics not included in the
NLO-FFN calculation, in addition to being well-behaved as $\mu ^2/m_H^2$
becomes large --- according to the renormalization group equation. Similar
comments apply to the other 2$\rightarrow $2 $\tilde{\sigma}$ terms and the
corresponding subtraction. The GF1 terms will be proportional to $(d_g^H-\,^1%
\tilde{d}_g^H\otimes d_H^H)$ where $d_{g,H}^H$ are the fully evolved and $%
\,^1\tilde{d}_g^H$ the perturbative fragmentation functions.

%% JCC
The following (terms in)
parton distribution and fragmentation functions $f_A^H$,
${}^1\tilde{f}_g^H$, ${}^1\tilde{f}_g^g$, and $d_g^H$ all vanish at
the `threshold', $\mu-m_H$
by {\em calculation} (cf.\ Eqs.~\ref{pertpdf},\ref{pertfrg} and Refs.~\cite
{ColTun,ACOT}). In addition, in the threshold region,
\begin{eqnarray}
d_H^H &\sim &\delta (1-z)+O(\alpha _s)  \nonumber \\
f_A^H-f_A^g\,\otimes \,^1\tilde{f}_g^H &\sim &O(\alpha _s^2)
\label{Threshold} \\
d_g^H-\,^1\tilde{d}_g^H\otimes d_H^H &\sim &O(\alpha _s^2)  \nonumber
\end{eqnarray}
Hence, the differences between the first two columns of the terms
represented in Fig.~\ref{fig:HadXsecB} --- (2$\rightarrow $2~$\tilde{\sigma}%
)- $heavy quark mass subtraction --- vanish even faster than the individual
terms approaching the threshold. Using these results, we obtain, in this
limit
\begin{equation}
\begin{array}{rclrr}
\sigma _{AB}^{HX} & \stackrel{\rm threshold}{\longrightarrow } &
\;\;f_{AB}^{l\,l^{\prime }}\,\otimes \,^2\tilde{\sigma}_{l\,l^{\prime }}^{H%
\bar{H}} & \hspace{0.16in}: & \mbox{LO-FFN} \\
&  & +\;f_{AB}^{l\,l^{\prime }}\,\otimes \,(\,^3\tilde{\sigma}_{l\,l^{\prime
}\;{\rm (virtual)}}^{H\bar{H}}+\,^3\tilde{\sigma}_{l\,l^{\prime }\;{\rm %
(real)}}^{H\bar{H}l^{\prime \prime }}) & : & \mbox{NLO-FFN} \\
& = & \;\;\tilde{\sigma}_{AB}^{HX}\;\;{\rm (full\;NLO-FFN\;scheme)} &  &
\end{array}
\label{FFNlimit}
\end{equation}
That is, the hadronic cross section in this formalism reduces to the full
flavor creation (HC) result of the FFN scheme to order $\alpha _s^3$. In
this region, there is effectively only one large momentum scale ($p_T\sim
m_H $); and the FFN scheme is well suited to represent the correct physics.

We see, therefore, the ACOT formalism provides a natural generalization of
the familiar light-parton pQCD to the case including quarks with non-zero
mass which contains the right physics over the entire energy range. At high
energies, Eq.~\ref{HadXsec1} gives the most natural description of the
underlying physics. The mass subtraction terms appearing in Eq.~\ref
{nloIRSxsec} for the $\alpha _s^3$ hard cross section correspond to the 
$\epsilon^{-1}$ poles arising from collinear singularities in the
$m_H=0$ QCD
parton formalism which are usually removed by $\overline{{\rm MS}}$
regularization.\footnote{%
\tightenlines
Consistency of this generalized formalism with the usual $\overline{{\rm MS}}
$ scheme at high energies is ensured by adopting the precise definitions of 
$\alpha_s(\mu)$ and $f_A^{g,H}$, $d_{G,H}^H$ in the renormalization scheme of
Ref.~\cite{ColTun,CWZ}. As mentioned earlier, in this scheme, the parton
distributions and fragmentation functions satisfy the familiar $\overline{%
{\rm MS}}$ evolution equations above the respective thresholds.} On the
other hand, at energy scales comparable to the quark mass $m_H$, consistency
with the physically sensible NLO-FFN calculation in the threshold region is
also guaranteed by keeping the full $m_H$ dependence in the partonic cross
sections appearing in Fig.~\ref{fig:HadXsecB}, as described in the previous
paragraph.

\section{Calculations}

\label{sec:Ingredients}

Our numerical calculations are carried out using Eq.~\ref{HadXsec1}, with
hard cross sections given by Eqs.~\ref{loPxsec} and \ref{nloIRSxsec}. (The
%%JCC
equations are presented graphically in Fig.~\ref{fig:HadXsecB}.) We briefly
describe how the various quantities on the right-hand-side of the equation
are obtained.

For the parton distributions $f_A^a(x,Q),$ we use the CTEQ3M set \cite{CTEQ}
for definiteness. The CTEQ parton distribution sets, in general, are evolved
according to the ACOT scheme described in Ref.~\cite
{ACOT,ColTun,CWZ,Collins97} which is the one we use in the heavy quark
production theory. This is necessary for obtaining consistent results
(a point which has not been entirely obvious to all users of the scheme).
The reason is: the expected compensation in the threshold region between the
QCD-evolved $f_A^H(x,Q)$ and the perturbatively generated subtraction $%
f_A^H(x,Q)\,_{SUB}\equiv f_A^g(x,Q)\otimes \,^1\tilde{f}_g^H(x,Q)$ (cf.\ Eq.~%
\ref{MassSubtr}) will not take place unless the choice of evolution scheme
and the choice of the location of the heavy quark threshold match properly.%
\footnote{%
\tightenlines
%% JCC
As an example, MRS distributions use $\mu=2m_H$ as the threshold for evolving 
$f^H(x,\mu ),$ from zero in contrast to $\mu=m_H$ which is
required by our renormalization
scheme. Using MRS distributions in this framework represents a mismatch of
schemes, and will lead to unphysical results as described below.
If a matching point other than $\mu=m_H$ is used, then the heavy quark
distribution must start from a non-zero value.
} This is
%%
%% JCC.  I have changed Q to \mu where appropriate.
illustrated in Fig.~\ref{fig:Compensate} which compares $f_A^H(x,\mu )$ and 
$f_A^H(x,\mu)\,_{SUB}$ for charm and bottom as a function of $\mu$ at $x=0.1$ 
and
$x=0.01$. Each curve individually vanishes at the threshold $\mu =m_H$ as
required in this general scheme. As $\mu $ increases, both grow at a rapid rate
because the evolution is driven by the large gluon distribution (through the
mass-independent splitting kernel); however, the difference of the two,
which determines the actual correction to the main contribution in this
region (due to the HC process), grows slowly as one would expect on physical
grounds. A failure to ensure the proper compensation between these terms due
to a mismatch of schemes or the location of the threshold can lead to quite
unphysical results because then the difference would be of the same order of
magnitude as the individual terms.\footnote{%
\tightenlines
$f_A^H(x,\mu )$ and $f_A^H(x,\mu)\,_{SUB}$ 
are not expected to cancel at large $\mu$; the latter, in
fact contains the divergent $\ln (\mu /m_H)$ factor in that limit. In that
region, this divergent subtraction term plays the other important role of
cancelling the corresponding large logarithm in the NLO-FC term to render
the latter infra-red safe, as shown in Eq.~\ref{nloIRSxsec}.}
\figCompensate

In calculating the partonic cross sections which appear in Eqs.~\ref{loPxsec}
and \ref{nloIRSxsec}, the formula for the LO-FFN cross sections $^2\tilde{%
\sigma}_{gg}^{HH}$ and $^2\tilde{\sigma}_{gH}^{gH}$ are well-known. For the
NLO-FFN cross section, both virtual ${}^3\tilde{\sigma}_{gg}^{HH}$ and real 
${}^3\tilde{\sigma}_{ll^{\prime }}^{HHl^{\prime \prime }}$, we used the
Fortran codes from Ref.~\cite{Smithetal} and Ref.~\cite{MNR}.

To calculate the gluon and heavy-quark fragmentation functions appearing in
Eq.~\ref{HadXsec1}, we need also the QCD evolved fragmentation functions $%
d_{g,H}^H$. As mentioned earlier, for the purpose of this paper, we restrict
ourselves to the $H$-quark production cross section; hence $d_{g,H}^H$ are
parton to parton fragmentation functions. To obtain heavy flavor hadron
production cross sections, it is necessary to perform an additional
convolution with the appropriate hadronic fragmentation functions. The QCD
evolved fragmentation functions $d_{g,H}^H(z,\mu )$ are generated by solving
numerically the QCD evolution equation in our scheme, using as input at $%
\mu_0=m_H$ the perturbative formula, cf.\ Ref.~\cite{MelNas91,CacGre}. The
comments about proper compensation between parton distributions made above
also apply to the fragmentation functions $d_{g,H}^H$ and $^1\tilde{d}%
_{g,H}^H$. These features have been examined in detail during our
calculation.
%TCIDATA{LaTeXparent=0,0,hhk1.TEX}

%TCIDATA{ChildDefaults=chapter:4,page:1}

%% WKT 01 Oct 97

\section{Results}

\label{sec:Results}We now present typical results for $b$ quark production
cross section $\frac{d\sigma }{dp_t^2dy}|_{y=0}$ {\it vs}.\ $p_t$ and {\it vs%
}.\ the QCD scale parameter $\mu $ at collider energies. For simplicity, as
is customary in the literature, we use a single scale parameter $\mu $ to
represent the renormalization scale, the factorization scale for the parton
distributions, and the factorization scale for the fragmentation functions.
In principle, these could be chosen as independent; the hard cross sections
would then depend on all three scales. As a rule, we shall express the scale
$\mu $ as multiples of the natural physical scale $M_T\equiv \sqrt{%
p_t^2+m_H^2},$ although, again, other choices could also be considered.
Except for explicit discussions concerning the $\mu $-dependence of the
cross sections, our default choice of scale is $\mu ^2=M_T^2/2$. In order to
be able to clearly discern the various contributions to the steeply falling
function $\frac{d\sigma }{dp_t^2dy}(p_t)$, we shall in general use the
scaled cross section $\ddot{\sigma}(p_t)\equiv p_t^5\frac{d\sigma }{dp_t^2dy}%
|_{y=0}$ when examining its $p_t$ behavior.\footnote{%
\tightenlines
We also note, in evaluating the right-hand-sides of
Eq.~\ref{HadXsec1} (cf.\
also Fig.\ \ref{fig:HadXsecB}), we have uniformly omitted the last convolution
with $d_H^H$ for simplicity of calculation. The numerical effect of
including this factor is relatively small.} We concentrate mostly on
b-production at the Tevatron for definiteness.

\subsection{Inclusive $p_t$ distribution and comparison of
heavy parton picture in the FFN scheme}

Fig.~\ref{fig:Figx1}a shows $\ddot{\sigma}(p_t)$ vs.\ $p_t$ for $b$
production at 1800 GeV, including the individual terms on the
right-hand-side of Eq.~\ref{HadXsec1}.\footnote{%
\tightenlines
In terms of Fig.~\ref{fig:HadXsecB}, the mass-subtraction terms are combined
with the associated order $\alpha _s^3$ HC1 terms (from which they
originate) to yield the infra-red safe hard cross-sections $\hat{\sigma}$.}
Over the range $10<p_t<100$ GeV, the two largest contributing terms are
``leading order'' ($2\rightarrow 2,$ tree-level) HE1 ($\hat{\sigma}%
_{gH\rightarrow gH}$) and HC0 ($\hat{\sigma}_{gg\rightarrow H\bar{H}}$); the
other tree-level terms and the ``next-to-leading'' term, HC1 ($\hat{\sigma}%
_{gg\rightarrow gH\bar{H}},$ mass-subtracted), constitute about $10-25\%$ of
the cross section, depending on the value of $p_t$. The left-hand-side plot
shows the contributions from the individual terms; and the right-hand-side
plot compares the relative sizes of the combined LO-VFN (i.e. tree-level)
contribution, the NLO-VFN contribution and the total cross-section TOT-VFN 
when the calculation is organized in this scheme.

%%%%%%%%%%%%%%%%%%%%%%%%%%%%%%%%%%%%%%
\figonex
%%%%%%%%%%%%%%%%%%%%%%%%%%%%%%%%%%%%%%

Two interesting features are worth noting. First, {\em the LO-VFN contributions
(tree processes) give a reasonable approximation to the full cross section},
the NLO-VFN correction is relatively small. (This is in sharp contrast to the
situation in the familiar FFN scheme where the NLO-FFN term is bigger than the
LO-FFN one. Cf.\ Fig.~\ref{fig:Figx2} and discussions below.) This is, 
of course,
an encouraging result, suggesting that the heavy quark parton picture
represents an efficient way to organize the perturbative QCD series.
Secondly, {\em the HE1 contribution is comparable to, and even somewhat
larger than, the HC\O\ one} -- in spite of the smaller heavy quark parton
distribution in the initial state compared to the gluon distribution. Closer
examination reveals that two effects contribute to this non-apparent result:
a larger color factor for the HE1 process, and the presence of
$t$-channel
gluon exchange diagrams which is absent in the HC\O\ process.\footnote{%
\tightenlines
The precise values of the HE1 contribution are somewhat sensitive to the
choice of factorization scheme and scale, especially close to the threshold
region, as will be shown below. However, one of the important features of
our formalism is that any scheme and scale dependence in HE1 will be
closely matched by changes in the HC1 contribution (through the
corresponding subtraction term), so that the combined inclusive cross
section remains relatively stable. {\it Cf}.\ discussions in the previous
section about the matching of evolved and perturbative parton distributions.
For the current discussion, we adopt $\mu =M_T/\sqrt{{2}}$ as a central
choice, given commonly used ranges of $\mu $ such as $[M_T/2,2M_T]$ and $%
[M_T/4,M_T]$.}

%%%%%%%%%%%%%%%%%%%%%%%%%%%%%%%%%%%%%%
\figtwox
%%%%%%%%%%%%%%%%%%%%%%%%%%%%%%%%%%%%%%

It is useful to compare the above situation with the same results organized
in a way more familiar from the conventional FFN scheme point of view. For
this purpose, one begins with the intermediate cross-sections $\tilde{\sigma}
$ for heavy flavor creation processes only, HC0 ($\tilde{\sigma}%
_{gg\rightarrow H\bar{H}}$) and HC1 ($\tilde{\sigma}_{gg\rightarrow gH\bar{H}%
},$ no mass-subtraction). {\em Corrections to the FFN scheme calculations}
in the full scheme then consist of the remaining terms on the
right-hand-side of the cross section formula depicted in
Fig.~\ref{fig:HadXsecB}%
, which are now most naturally organized with the mass-subtraction terms
combined with the corresponding $2\rightarrow 2$ cross-sections in the same
row. Fig.~\ref{fig:Figx2}a,b show $\ddot{\sigma}(p_t)$ {\it vs}.\ $p_t$ in
the same format as in the previous plot but with individual contributions
organized in this way.\footnote{%
\tightenlines
 For the purposes of this comparison, we use the same 5-flavor PDF's for both
Fig.~\ref{fig:Figx1}a,b and Fig.~\ref{fig:Figx2}a,b so that the only 
difference is how we combine the terms. 
Using 4-flavor PDF's (as would be appropriate for the FFN scheme, without
the subtraction terms) yields 
virtually indistinguishable curves differing by 
$\sim 1$\% at $p_t = 10$ GeV, 
and $\sim 3$\% at $p_t = 100$ GeV.
}
 The largest term is now the ``NLO'' HC1 ($\tilde{%
\sigma}_{gg\rightarrow gH\bar{H}}$) followed by the ``LO'' HC0 of the
conventional FFN scheme. The fact that {\em the NLO }(order $\alpha _s^3$)%
{\em \ term }$\tilde{\sigma}_{HC1}$ {\em is much larger than the LO term }%
(order $\alpha _s^2$) $\tilde{\sigma}_{HC0}$ -- the ``K-factor'' is
typically of the order $\sim 2.5$ -- is disturbing from the perturbation
theory point of view, as has been known since the order $\alpha _s^3$
calculations were first done. On the other hand, we see from Fig.~\ref
{fig:Figx2}a, {\em the corrections to the FFN scheme terms}, consisting of
the other terms in Fig.~\ref{fig:HadXsecB},
{\em are positive but not very large}
-- again of the order of $\lsim$20\%. This means that the effects of
resumming the collinear logarithms, represented by these additional terms,
are modest for this case -- a non-obvious result on account of the large
K-factor and the significant $\mu $-dependence of the FFN scheme
calculations (see next subsection). The net effect of these correction terms
is to increase the theoretical cross-section. This is encouraging since the
NLO FFN scheme result is known to be systematically smaller than the
experimentally measured cross section at the Tevatron. However, this
increase appears to fall short of the current observed 
discrepancy~\cite{FMNR97a,Berger}.
Cf.\ Fig.~\ref{fig:b_expt}.
\figbexpt

A comparison of Figs.~\ref{fig:Figx1} and \ref{fig:Figx2} shows that,
interestingly, the HE1 ($\hat{\sigma}_{HE1}$) contribution to the heavy
quark production cross section in the heavy quark parton picture is quite
comparable to the HC1 contribution in the complementary FFN scheme view 
($\tilde{\sigma}_{HC1},$ no mass-subtraction) -- to within about
10\%. Thus, at least for this energy range, {\em the heavy quark parton
picture overlaps considerably with the FFN heavy flavor creation picture as
far as the inclusive }$p_t${\em \ distribution is concerned}\footnote{%
\tightenlines
An equivalent way of saying this is: the subtraction terms, which represent
the overlap between the two, are a reasonable approximation to both in this
energy range. Thus the ``correction'' to either one, represented by the
combination of the other with the corresponding subtraction, are relatively
small---as demonstrated above.}{\em \ }--- these two pictures are
complementary rather than mutually exclusive, as sometimes perceived in the
literature. It is, of course, much easier to calculate the tree-level HE1
cross section (a text book case) than the HC1 one (a tour de force).
Thus, for this physical quantity, the heavy quark parton picture represents
a much more efficient way to arrive at the right answer. This approximate
equivalence between the HE1 and HC1 contributions to the {\em inclusive}
$p_t$ cross section cannot, of course, be taken literally.\footnote{%
\tightenlines
The HC1 diagram, of course, contains a lot more information on detailed
differential distributions (such as non-back-to-back-jets) 
which is not contained in
the HE1 diagram.} The two contributions do not have the same ($s,\,p_t,\mu
,y$) dependence -- in fact, the $\mu $ dependence can be rather different,
as we will discuss next.

\subsection{Scale dependence of the cross section}

In Fig.~\ref{fig:Figx4} we show a representative plot of $\sigma (p_t,\mu )$
vs.\ $\mu $ at $p_t=20$ GeV. The tree-level HC0 and HE1 terms give the
dominant contributions. In addition to the common $\alpha _s^2(\mu )$
factor, HC0 is predominantly driven by the gluon distribution, and this is a
decreasing function of $\mu $. On the other hand, the tree-level HE1 term is
driven by the heavy-quark distribution, and this is a increasing function of
$\mu $. These two components compensate each other. Thus the full LO
cross-section in the VFN ACOT scheme has a moderate $\mu$ dependence.

%%%%%%%%%%%%%%%%%%%%%%%%%%%%%%%%%%%%%%
\figfourx
%%%%%%%%%%%%%%%%%%%%%%%%%%%%%%%%%%%%%%

The situation is different for the FFN scheme, shown in Fig.~\ref{fig:Figx5}.
Here one finds that both the LO-FFN and NLO-FFN results (proportional to
light parton distributions) are decreasing functions of $\mu $, resulting in
a steep $\mu $ dependence for the combined result, TOT-FFN. If we were to
compute higher order corrections in the FFN scheme, we would eventually
observe compensating terms to reduce the $\mu $ dependence (as we know the
``all-orders'' result must be independent of $\mu $). The corrections to the
FFN scheme result which have been resummed in the ACOT scheme into HE and GF
contributions (minus subtractions) represent a part of these higher-order
effect.

%%%%%%%%%%%%%%%%%%%%%%%%%%%%%%%%%%%%%%
\figfivex
%%%%%%%%%%%%%%%%%%%%%%%%%%%%%%%%%%%%%%

In Fig.~\ref{fig:Figx6} we compare directly the TOT-FFN and the TOT-VFN
results. The TOT-VFN result ranges from $\sim 5\%$ to $\sim 25\%$ above the
TOT-FFN result depending on the choice of $\mu $. The additional resummed
terms (labeled {\em Difference}) is seen to improve the $\mu $ dependence.
(See also Fig.~\ref{fig:Figx12b14tev} below for LHC energies.) \ Perhaps the
improvement is not as complete as one might expect. This suggests there is
still some non-negligible physics missing from the calculation. One possible
source might be the NLO-HE process (HE2), on account of the large size of
HE1. This is an ${\cal O}(\alpha _s^3)$ contribution which has yet to be
computed.

%%%%%%%%%%%%%%%%%%%%%%%%%%%%%%%%%%%%%%
\figsixx
%%%%%%%%%%%%%%%%%%%%%%%%%%%%%%%%%%%%%%

Fig.~\ref{fig:Figx12b1800} displays the scaled cross section vs. the
physical variable \ $p_T$ for the range of $\mu $ $[M_T/2,2M_T]$. The upper
band corresponds to $\mu =M_T/2$, and the lower to $\mu =2M_T$. We see the
increase in central value of the cross-section as well as the reduction in
scale dependence in the ACOT scheme compared to the FFN scheme. The
improvement is not dramatic in either case at this energy.
%%%%%%%%%%%%%%%%%%%%%%%%%%%%%%%%%%%%%%
\figtweax
%%%%%%%%%%%%%%%%%%%%%%%%%%%%%%%%%%%%%%
One expects this to change at higher energies. Fig.~\ref{fig:Figx12b14tev}
shows the corresponding results at the LHC energy of $\sqrt{s}=14\ TeV$.
For comparison, we also present results for b-production at the CERN $Sp\bar{%
p}S$ energy of $\sqrt{s}=630\ GeV$ in Fig.~\ref{fig:Figx12b630}.

%%%%%%%%%%%%%%%%%%%%%%%%%%%%%%%%%%%%%%
\figtwebx
%%%%%%%%%%%%%%%%%%%%%%%%%%%%%%%%%%%%%%
%%%%%%%%%%%%%%%%%%%%%%%%%%%%%%%%%%%%%%
\figtwecx
%%%%%%%%%%%%%%%%%%%%%%%%%%%%%%%%%%%%%%

\subsection{The Rapidity Distribution}

Although the inclusive HE1 contribution in the ACOT was roughly comparable 
to the HC1 term in the FFN schemes, the dependence
on the individual variables ($s,\,p_t,\mu ,y$) can be quite different.
Having already investigated the $\mu $-dependence, we turn to the rapidity
dependence of the underlying processes.

In Fig.~\ref{fig:Figx8} we compare the rapidity distribution for the HE1
and the HC\O\  processes. To more easily compare the relative shape, we have
scaled the two curves to equal area. We observe that the HE1 process
yields a broader rapidity distribution than the HC\O\  processes. In part,
this is expected as the HE1 process includes a t-channel gluon exchange
which can give an enhanced contribution in the forward direction. To see the
relative effect on the rapidity distribution for the complete
next-to-leading order calculations, in Fig.~\ref{fig:Figx9} we display the
ratio of the TOT-VFN compared to the TOT-FFN cross section as a function of
the rapidity.%%%%%%%%%%%%%%%%%%%%%%%%%%%%%%%%%%%%%%
\figeightx
%%%%%%%%%%%%%%%%%%%%%%%%%%%%%%%%%%%%%%
%%%%%%%%%%%%%%%%%%%%%%%%%%%%%%%%%%%%%%
\figninex
%%%%%%%%%%%%%%%%%%%%%%%%%%%%%%%%%%%%%%

\subsection{Comment on related work}

It is worth mentioning that the resummation of large logarithms in the
fixed-order calculations into parton distributions and fragmentation
functions has also been studied by Cacciari and Greco \cite{CacGre}. In the
notation of Eq.~\ref{FacThmH}, their approach amounts to adopting the
following ansatz for the hadro-production cross section:
\begin{equation}
\sigma _{AB}^{HX}=\sum_{a,b,c}f_{\left\{ AB\right\} }^{\left\{ ab\right\}
}\otimes \hat{\sigma}_{ab}^{c,r}(m_H=0)\otimes d_c^H  \label{CacGre}
\end{equation}
where ($a,b,c$) are summed over all parton flavors including $H,$ and $\hat{%
\sigma}_{ab}^{c,r}(m_H=0)$ is given by the zero-mass (i.e.\ light-parton)
NLO jet cross section calculation. This is a good approximation in the
asymptotic region $p_T\gg m_H,$ but it does not reproduce the right physics
when $p_T$ is not significantly larger than the quark mass. The main result
of their calculation was that the predicted $b$-production cross section has
less scale dependence than the FFN scheme result, but it lies within the
uncertainty band of the latter. Hence their predictions lie substantially
below the experimental measurement and somewhat below our results. Our
theory resums the same large logarithms associated with final state
collinear singularities. But in addition, (i) we treat initial state parton
distributions and final state fragmentations of the heavy quark
symmetrically; and, more importantly, (ii) by keeping the heavy quark mass
in the hard cross section according to the general factorization theorem,
our results are applicable over the entire energy range. Strictly speaking,
by using the order $\alpha _s^3$ jet cross section in Eq.~\ref{CacGre}, Ref.~%
\cite{CacGre} includes higher order corrections to the corresponding terms
in our calculation (GF1 and HF1 terms in Eq.~\ref{HadXsec1}). However, as
shown above, the contribution from the latter is already small. Hence, the
corrections are not expected to be important, and they are, in any case, of
the same numerical order as NNLO flavor creation terms which are not
currently calculated. Nonetheless, this difference may be responsible for
the seemingly smaller $\mu $ dependence of their results. It is hard to be
sure because their calculation is applicable only when $p_T\gg m_H$.

%TCIDATA{LaTeXparent=0,0,hhk1.TEX}
                      
%TCIDATA{ChildDefaults=chapter:6,page:1}

%% WKT 01 Oct 97

\section{Concluding Remarks}

In this paper, we have systematically developed the theory for
hadro-production of heavy quarks according to the natural pQCD scheme of
Ref.~\cite{ColTun,AOT,ACOT,Collins97} which generalizes the conventional
(zero-mass) ``improved QCD parton model'' to include quark mass effects.
This formalism has the advantage that it contains the correct physics over a
wide range of energies and $Q$ ($p_T$), and it reduces to the conventional
results both at the low and the high energy limits. In particular, it
coincides with the widely used FFN scheme in its natural region of
applicability where $Q$ is or the same order of magnitude as $m_H$ -- the
one-large-scale region. Improvement over the FFN scheme results become
important when $Q\gg m_H$. This is manifested both in the reduced
theoretical uncertainty and in the increased cross-section. The improvement
comes at the price of somewhat more complicated calculations. As seen in Eq.%
\ref{nloIRSxsec} and Fig.\ref{fig:HadXsecB}, in addition to the FFN scheme
contributions, one needs to compute the other terms involving flavor
excitation and fragmentation processes and their subtractions. As emphasized
in Sec.~\ref{sec:Ingredients}, these calculations must be done with care,
due to the delicate cancellations required.

This more complete and consistent theory is, however, not a cure-for-all.
Within our scheme, the residue scale dependence seen in Sec.\ref{sec:Results}
may suggest that certain non-negligible higher order terms still need to be
included.  Our cross-section predictions, although higher than the existing
FFN ones, still fall somewhat short of the $b$ cross-section measured at the
Tevatron.  Some physics effects not included in our formalism could be
important: notably, those related to large logarithms of the type $\ln
(Q^2/s),\ln (m_H^2/s)$ $\sim \ln x$ which needs to be separately resummed.
This is another example of the {\em small-x problem}.\cite{Smallx} In this
paper, we also have not addressed questions concerning the hadronization of
the heavy quark which is not yet fully understood. 

As an important physical process involving the interplay of several large
scales, heavy quark production poses a significant challenge for further
development of QCD theory. 

\section*{Acknowledgments}

We would like to thank J.C.~Collins,
R.K.~Ellis, B.~Harris, M.L.~Mangano, P.~Nason,
J.~Smith, and D.E.~Soper for useful discussions. We are indebted to
B.~Harris and J.~Smith, and to M.L.~Mangano, P.~Nason, and G.~Ridolfi for
the use of their FORTRAN code in calculating the order $\alpha _s^3$
fixed-flavor-number scheme cross sections. 
 FIO thanks the Fermilab Theory Group for their kind hospitality during the 
period in which this research was carried out. 
 This work is partially supported
by the National Science Foundation, the U.S. Department of Energy, and the
Lightner-Sams Foundation.

\end{document}